\documentstyle[preprint,aps,prl,psfig]{revtex}
\tighten
\draft
\begin{document}

\draft

\preprint{\today}
\title{Static and Dynamic Properties of a Viscous Silica Melt}
\author{J\"urgen Horbach and Walter Kob}
\address{Institut f\"ur Physik, Johannes Gutenberg-Universit\"at,
Staudinger Weg 7, D-55099 Mainz, Germany}
\maketitle

\begin{abstract}
We present the results of a large scale molecular dynamics computer
simulation in which we investigated the static and dynamic properties
of a silica melt in the temperature range in which the viscosity of the
system changes from $O(10^{-2})$~Poise to $O(10^{2})$~Poise.  We show
that even at temperatures as high as 4000~K the structure of this
system is very similar to the random tetrahedral network found in
silica at lower temperatures. The temperature dependence of the
concentration of the defects in this network shows an Arrhenius
law. From the partial structure factors we calculate the neutron
scattering function and find that it agrees very well with experimental
neutron scattering data. At low temperatures the temperature dependence
of the diffusion constants $D$ shows an Arrhenius law with activation
energies which are in very good agreement with the experimental values.
With increasing temperature we find that this dependence shows a
cross-over to one which can be described well by a power-law, $D\propto
(T-T_c)^{\gamma}$. The critical temperature $T_c$ is 3330~K and the
exponent $\gamma$ is close to 2.1. Since we find a similar cross-over
in the viscosity we have evidence that the relaxation dynamics of
the system changes from a flow-like motion of the particles, as
described by the ideal version of mode-coupling theory, to a hopping
like motion. We show that such a change of the transport mechanism is
also observed in the product of the diffusion constant and the life
time of a Si-O bond, or the space and time dependence of the van Hove
correlation functions.  
\end{abstract}


\section{Introduction}
\label{sec1}
In recent years the dynamics of supercooled liquids has been the focus
of many investigations and due to these efforts our understanding of
this dynamics has increased significantly. Although very recently new
interesting phenomena in this dynamics have been found, such as the
dynamical heterogeneities~\cite{sillescu98} or the aging dynamics of
the system at very low temperatures~\cite{age_rev}, which so far are
not understood, we have by now at least a fair understanding of the
relaxation dynamics at temperatures a bit above the experimental glass
transition temperature $T_g$. In particular it has been demonstrated
that the so-called mode-coupling theory (MCT)~\cite{mct}, a theory
which relates the slowing down of the dynamics upon cooling to
nonlinear feedback effects, is able to give a qualitative and
quantitative correct description of glass-formers in which the
interaction between particles is not too different from the one of a
hard-sphere system, such as colloidal systems~\cite{colloids}, the
molecular glass-former orthoterphenyl~\cite{otp}, and even systems in
which some hydrogen bonding is present, such as
glycerol~\cite{glycerol}. The main prediction of the theory is that
there exists a critical temperature $T_c$ at which the relaxation
dynamics of the system changes qualitatively in that an unexpectedly
fast increase of the local activation energy of the transport
coefficients, such as the viscosity or the inverse of the diffusion
constant, is observed.  Qualitatively such a temperature dependence can
readily be seen in the so-called Angell plot~\cite{angell85}, where the
logarithm of the viscosity is plotted versus $T_g/T$, in that the
so-called fragile glass-formers show, at around 1.1-1.3$T_g$, a sharp
upward bend. Because of the similarity between this bend and the
prediction of MCT, many investigations on the dynamics of glass-formers
have been done for these types of systems, so that the validity of the
theory could be checked. Apart from this, most fragile systems have
glass transition temperatures which are readily accessible in an
experiment, thus making it relatively simple to investigate the
dynamics at and above $T_g$. This is not the case for most {\it strong}
glass-formers, since their $T_g$ is often above 500~K, such as SiO$_2$,
B$_2$O$_3$, or GeO$_2$, to name a few.  Because of their high value of
$T_g$ the {\it microscopic} dynamics of these systems is understood in
much less detail than the one of fragile glass-formers. Sidebottom {\it
et al.}~\cite{sidebottom93} have investigated the dynamics of
B$_2$O$_3$ by means of photon correlation spectroscopy, and very
recently Wischnewski {\it et al.}~\cite{wischnewski98} have used
neutron scattering experiments to study the dynamics of SiO$_2$ also at
high temperatures. Despite these investigations our knowledge on the
dynamics of strong glass-formers is significantly less than the one of
fragile ones.

Valuable insight into the dynamics of supercooled liquids have also
been obtained through computer simulations. Whereas many of the earlier
studies~\cite{simulation_rev} have investigated the dynamics of strong
glass-formers, most of the more recent studies focused on simple liquid
like systems, such as soft spheres or Lennard-Jones systems, since in
the latter type of models the interaction between the particles is much
simpler than in the one of strong glass-formers and hence the system
can be probed on larger time windows. Very recently however, the
investigation of strong glass-formers, mainly silica, by means of
computer simulations has become more popular again. From such
simulations it is hoped that insight is gained into, e.g., the nature
of the so-called boson-peak~\cite{bp_sim,taraskin97,horbach98b}, a
dynamical feature at around 1THz whose origin is still a matter of
debate~\cite{wischnewski98,bp_papers}, or to investigate the structure
and dynamics at different pressures and temperatures on a microscopic
level~\cite{strong_sim}.

Despite the mentioned investigations we are however still lacking a
comparable detailed characterization of the relaxation dynamics of
strong glass-formers as it has been obtained for simple
liquids~\cite{ss_sim,roux89,kob_lj,kob95b} and water~\cite{sciortino}.
In the present work we therefore present the results of a molecular
dynamics computer simulation of silica, the prototype of a strong
glass-former, in which such an analysis has been done. These results
will allow us to gain more insight into the dynamics of this system, to
relate it to its structural properties and to compare these results
with the ones obtained for fragile systems. The rest of the paper is
organized as follows: In Sec.~\ref{sec2} we give the details of the
model and the simulation, in Sec.~\ref{sec3} we present the results,
and in Sec.~\ref{sec4} we summarize and discuss them.

\section{Model and Details of the Simulations}
\label{sec2}

As already mentioned in the Introduction, various aspects of the
dynamics of amorphous silica have already been investigated in the
past. In these simulations potentials with different levels of accuracy
have been used, most of which give a satisfactory description of the
{\it structural} (i.e.~static) properties of amorphous silica. Very
recently Hemmati and Angell have pointed out, however, that the
predictions of these various potentials regarding {\it dynamical}
properties, such as the diffusion constant at low temperatures, can
differ by orders of magnitudes~\cite{hemmati98}. For the investigation
of the dynamical behavior it is therefore important to choose the 
potential with particular care. One of the most reliable potentials
seems to be the one proposed by van Beest {\it et al.} (BKS) which
was developed by using a mixture of {\it ab initio} calculations and 
classical lattice dynamics simulations~\cite{beest90}. The BKS
potential is given by
\begin{equation}
\phi(r_{ij})=\frac{q_i q_j e^2}{r_{ij}}+A_{ij}\exp(-B_{ij}r_{ij})-
\frac{C_{ij}}{r_{ij}^6}\quad .
\label{eq1}
\end{equation}
Here $r_{ij}$ is the distance between ions $i$ and $j$, and the values
of the partial charges $q_i$ and the constants $A_{ij}$, $B_{ij}$, and
$C_{ij}$ can be found in Refs.~\cite{beest90,vollmayr96b}. As we have
already done in previous investigations of this
model~\cite{horbach98b,vollmayr96b,vollmayr96a,horbach96,horbach98a,horbach98c},
we have truncated and shifted the non-Coulombic part of this potential
at a cut-off radius of 5.5\AA. This modification of the potential has
the positive effect that at normal pressure the density of the glass is
around 2.3g/cm$^3$, a value that is very close to the one of real
silica glass, 2.2g/cm$^3$~\cite{bruckner70}. In previous work we have
found that the dynamics of strong glass formers show quite significant
finite size effects~\cite{horbach96}. In order to avoid them 
we used a system which is relatively large for simulation of
supercooled liquids, i.e. 8016 ions. During the
runs the size of the cubic simulation box was fixed to $L=48.37$\AA.
Thus the density of the system was kept constant at 2.37g/cm$^3$,
which leads to a pressure which is a bit above normal pressure
(discussed in more detail below).

The Coulombic part of the potential was evaluated by means of
Ewald-sums with a constant $\alpha L$=12.82 ~\cite{allen90} and by
using for the Fourier part of the Ewald-sum all $k$-vectors of
magnitude less than $k_c=2\pi\sqrt{51}/L$. A short calculation shows that,
although such a small value of $k_c$ is sufficient to calculate the
interaction between two ions to a satisfactory accuracy, it would be
insufficient to set the {\it absolute} error of the {\it total}
Coulombic energy below an acceptable level. However, it can be
shown~\cite{horbach_diss}, that this small value of $k_c$ leads just
to a constant shift of the energy scale and hence does not affect the
forces.

Using the potential given by Eq.~(\ref{eq1}) and the masses of 28.086u
and 15.9994u for the silicon and oxygen atoms, respectively, we
equilibrated the system by doing a $NVT$ simulation (temperature was
kept constant by means of a stochastic collision algorithm). At each
temperature investigated, we made sure that the length of these
equilibration runs exceeded the typical relaxation times of the
system.  The temperatures considered were 6100~K, 5200~K, 4700~K,
4300~K, 4000~K, 3760~K, 3580~K, 3400~K, 3250~K, 3100~K, 3000~K, 2900~K
and 2750~K. In order to improve the statistics of the results we
averaged at each temperature over two independent runs. After the
equilibration we started microcanonical runs by integrating the
equations of motion with the velocity form of the Verlet algorithm. The
time step used was 1.6fs, which is sufficiently small to guarantee a
negligible drift of the total energy of the system at high and
intermediate temperatures (6100K$ \geq T \geq 3100$K) over the whole
run. For the lowest temperatures ($T \leq 3000$K) the runs were so long
that the drift in the energy could no longer be neglected.  Therefore
we rescaled every one million time steps the velocity of the particles
to bring the system back to the energy at which it was started. This
frequency should be small enough to avoid any effect on the relaxation
dynamics. At the lowest temperature the length of the runs were 12
million time steps (and 13 millions for the equilibration), giving a
total simulation time of about 20~ns. The overall computational effort
(equilibration plus production) for this temperature was about 13 years
of single processor time on a CRAY-T3E (or 2.5 months on 64
processors).

\section{Results}
\label{sec3}

In order to understand the dynamical behavior of the system it is
useful to know something about its static properties since we will show
that the transport mechanism of the atoms is intimately related to the
network structure of the system. Therefore we present in the following
subsection some of the static properties before we discuss in the
subsequent subsection the dynamics of the system.

\subsection{Static properties}
\label{sec3.1}

\begin{figure}[h]
\psfig{file=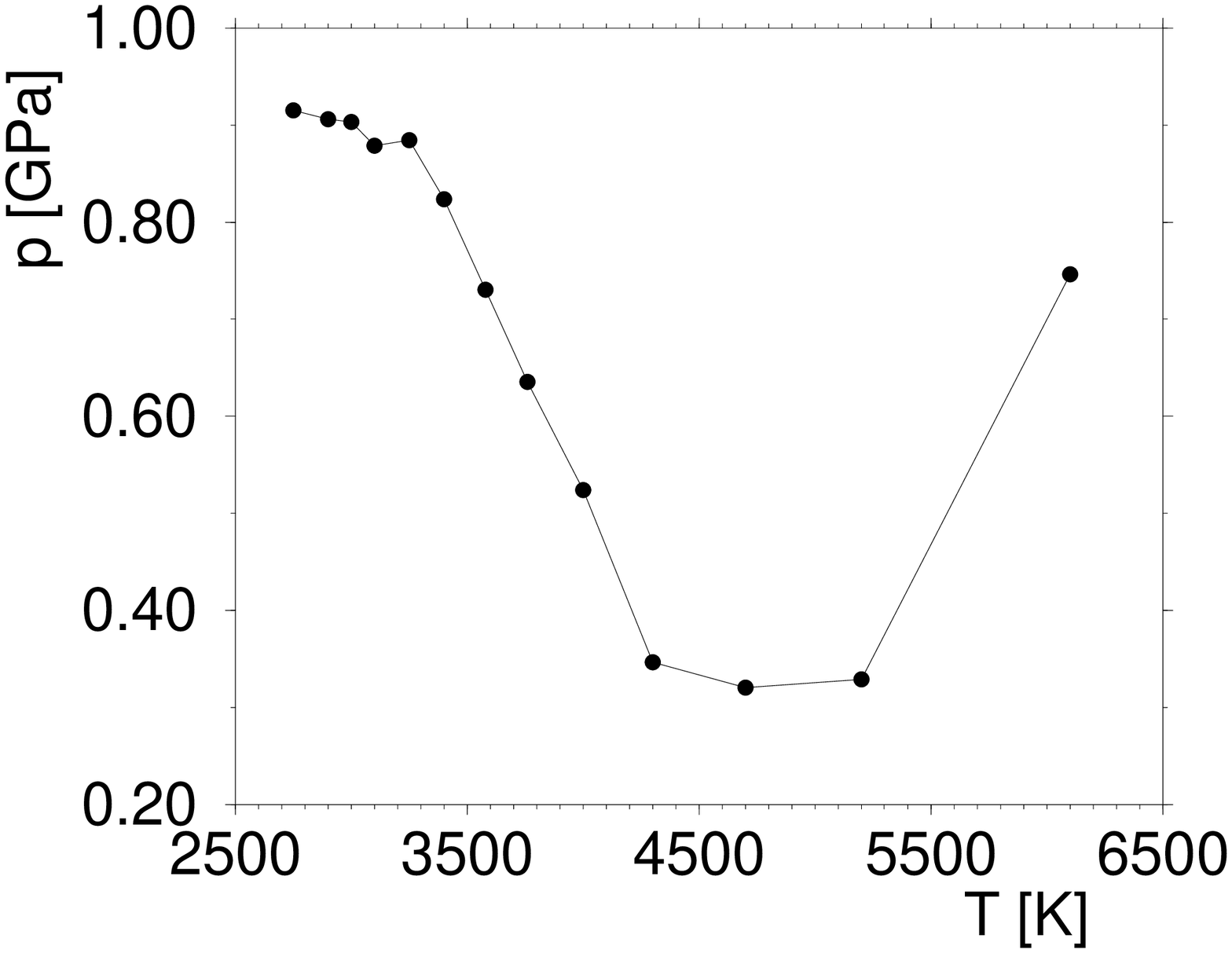,width=13cm,height=9.5cm}
\caption{Temperature dependence of the pressure.}
\label{fig1}
\end{figure}

One of the most remarkable properties of amorphous silica is that it
shows a density anomaly at around 1820~K~\cite{bruckner70}. Although in
a constant volume simulation such an anomaly can of course not be observed
in the density, it can easily be seen in the temperature dependence of
the pressure $p$, which is shown in Fig.~\ref{fig1}. We see that after
a fast decrease of $p$ between 6100~K and 5200~K, the pressure reaches
a minimum at around 4900~K, thus showing that the density at constant
pressure has a maximum, and then increases again with decreasing
temperature. Since the experimental value at which this density anomaly
occurs is 1820~K~\cite{bruckner70} we confirm the result of the
constant pressure simulation~\cite{vollmayr96b}, that the
BKS-potential overestimates the temperature at which the density
anomaly is observed.

\begin{figure}[h]
\psfig{file=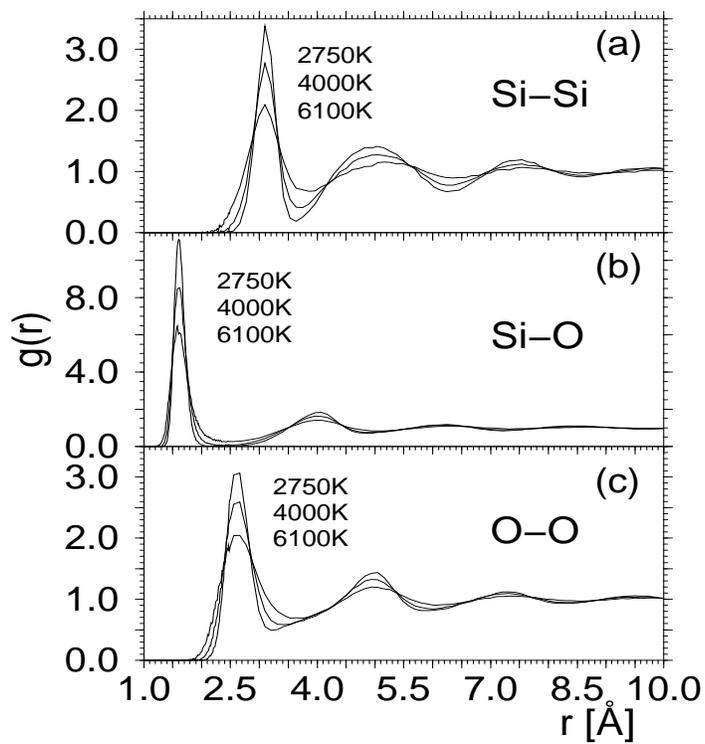,width=13cm,height=9.5cm}
\caption{Radial distribution functions for different temperatures. a)
Si-Si, b) Si-O, c) O-O.}
\label{fig2}
\end{figure}

That this extremum in pressure is not accompanied by a significant
change in the structural quantity $g_{\alpha\beta}$, the partial radial
distribution functions between particles of type $\alpha$ and
$\beta$~\cite{hansen86}, is demonstrated in Fig.~\ref{fig2}, where we
show the three $g_{\alpha\beta}$ for three different temperatures. What
can also be recognized from these figures is that the BKS-model
predicts that even at a temperature of 6100~K the arrangement of the
particles shows a structure which is much more similar to the one of a
liquid at its triple point than to the one of a gas. It should be
realized that this result is not in contradiction to the experimental
result that a silica melt at several thousand Kelvin does evaporate,
{\it if a free surface is present}. We have found that if the geometry
of the simulation does contain such a free surface the BKS model also
predicts that the atoms will evaporate from the system~\cite{roder99}.
However, due to the presence of the periodic boundary no free surface
is present in the simulation discussed in this work and thus no
evaporation takes place. This situation is thus similar to a melt in
the interior of the earth.

We also remark that in the temperature range considered, the time scale
of the relaxation dynamics of the system changes by about four decades
(this will be discussed in the next subsection). Thus we find that a
{\it relatively} small change in the structure of the melt is
accompanied by a dramatic change in the dynamics, an observation which
is qualitatively similar to the one found for simple
liquids~\cite{ss_sim,roux89,kob_lj,kob95b}. This fact thus gives
support to the point of view put forward by the mode-coupling
theory~\cite{mct}, which predicts that although the temperature
dependence of all structural quantities of a supercooled melt is very
weak, most dynamical quantities connected to the density fluctuations
of the system will show a very strong temperature dependence. Below we
will discuss these aspects in more detail.

\begin{figure}[h]
\psfig{file=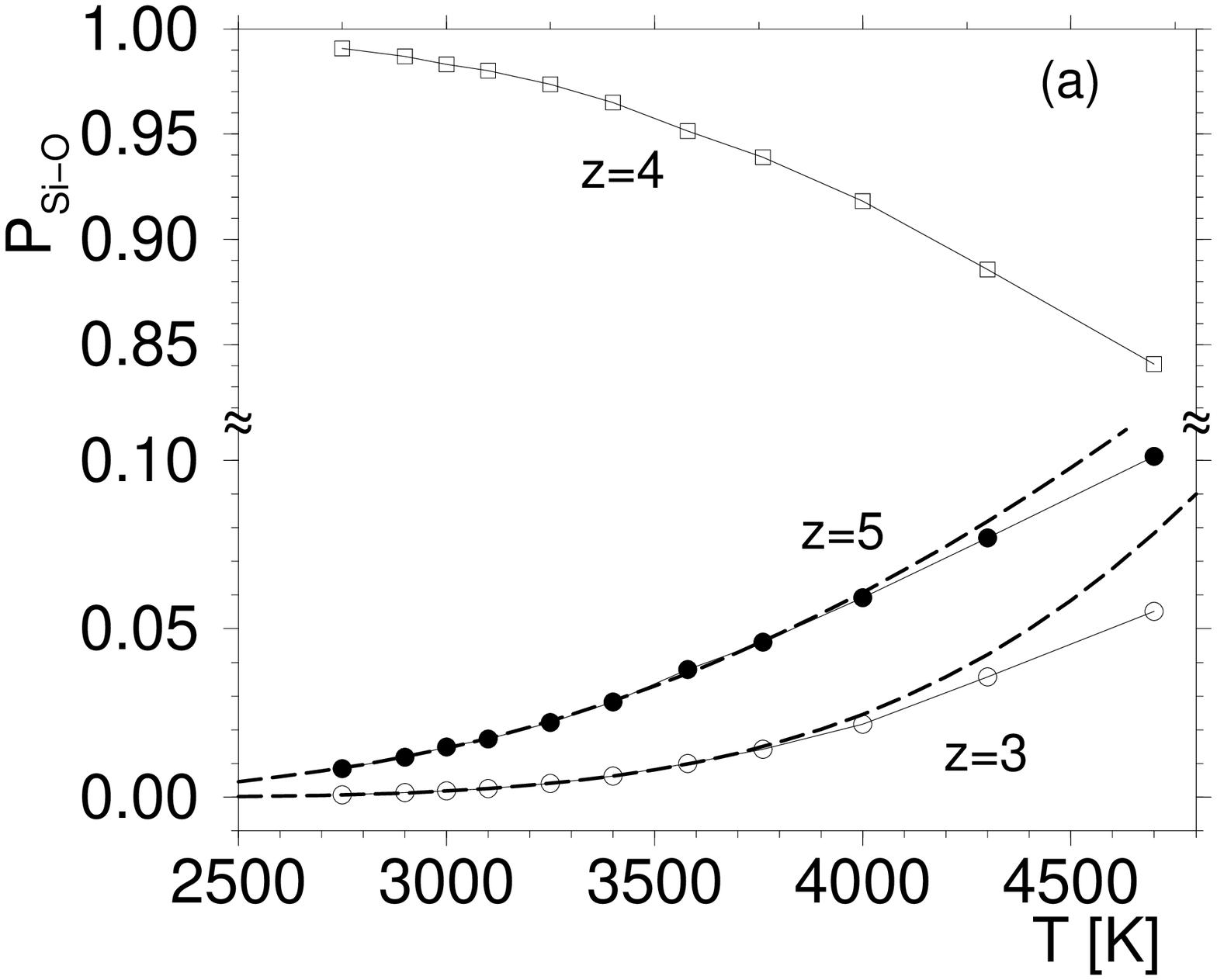,width=13cm,height=9.5cm}
\psfig{file=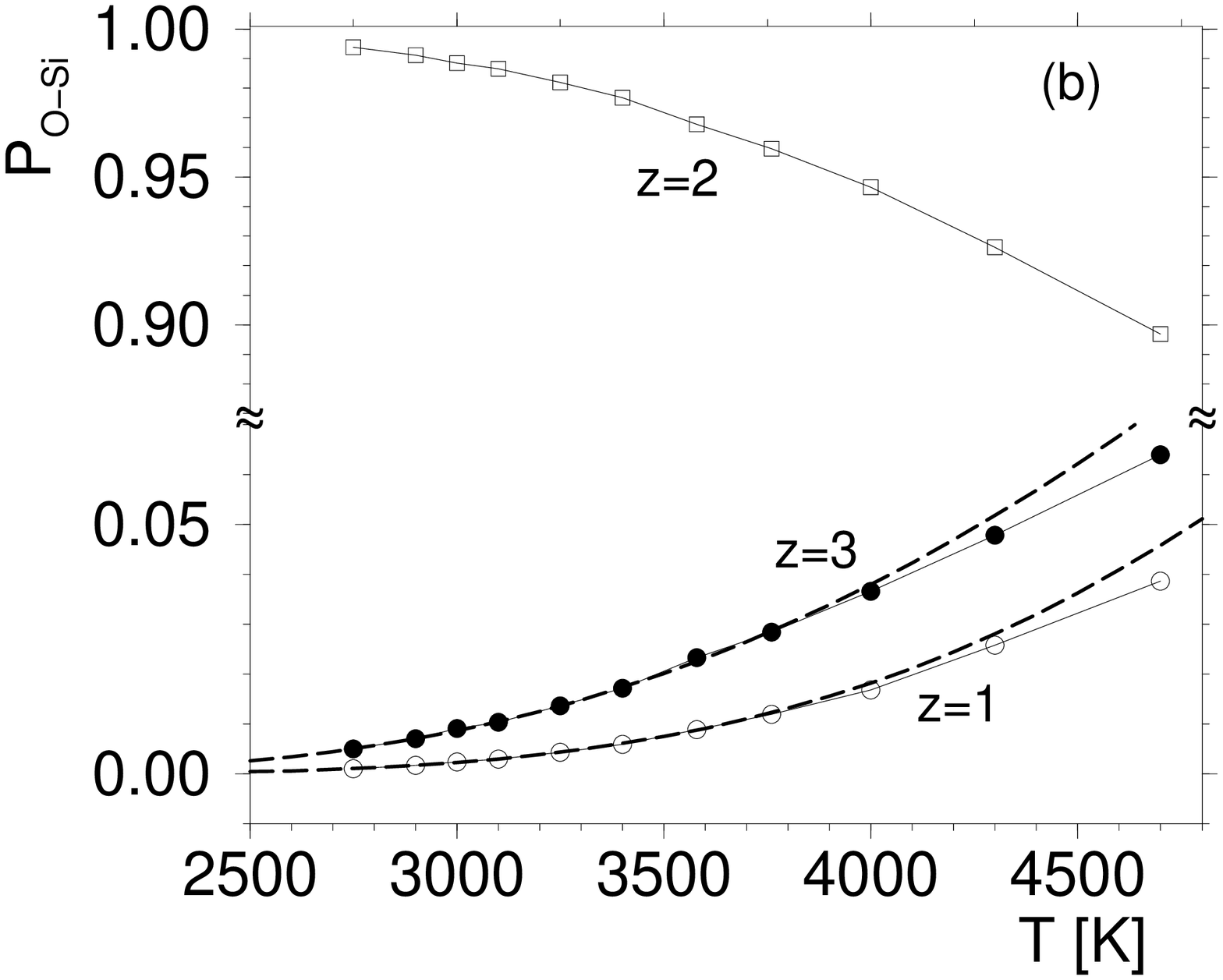,width=13cm,height=9.5cm}
\psfig{file=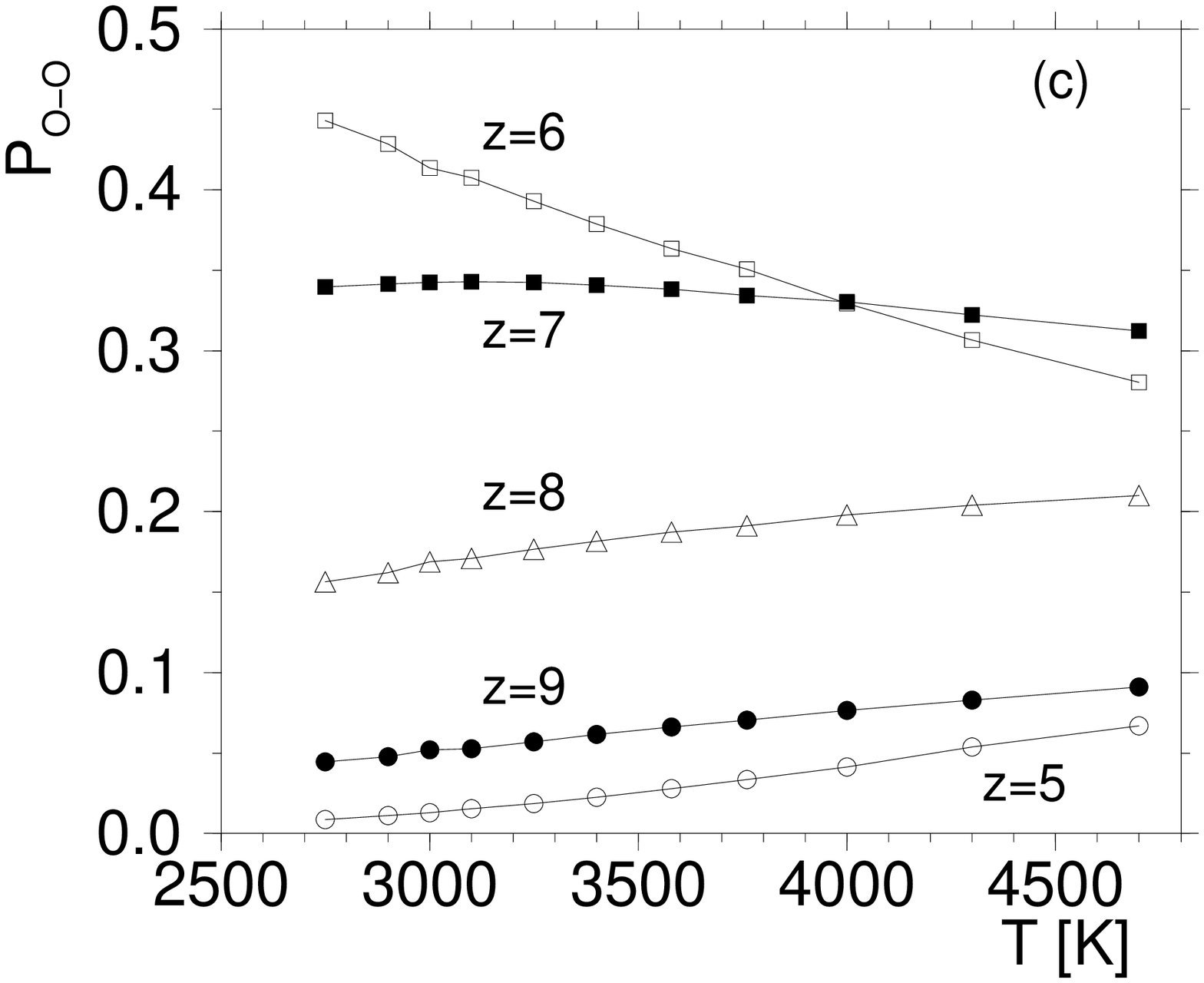,width=13cm,height=9.5cm}
\psfig{file=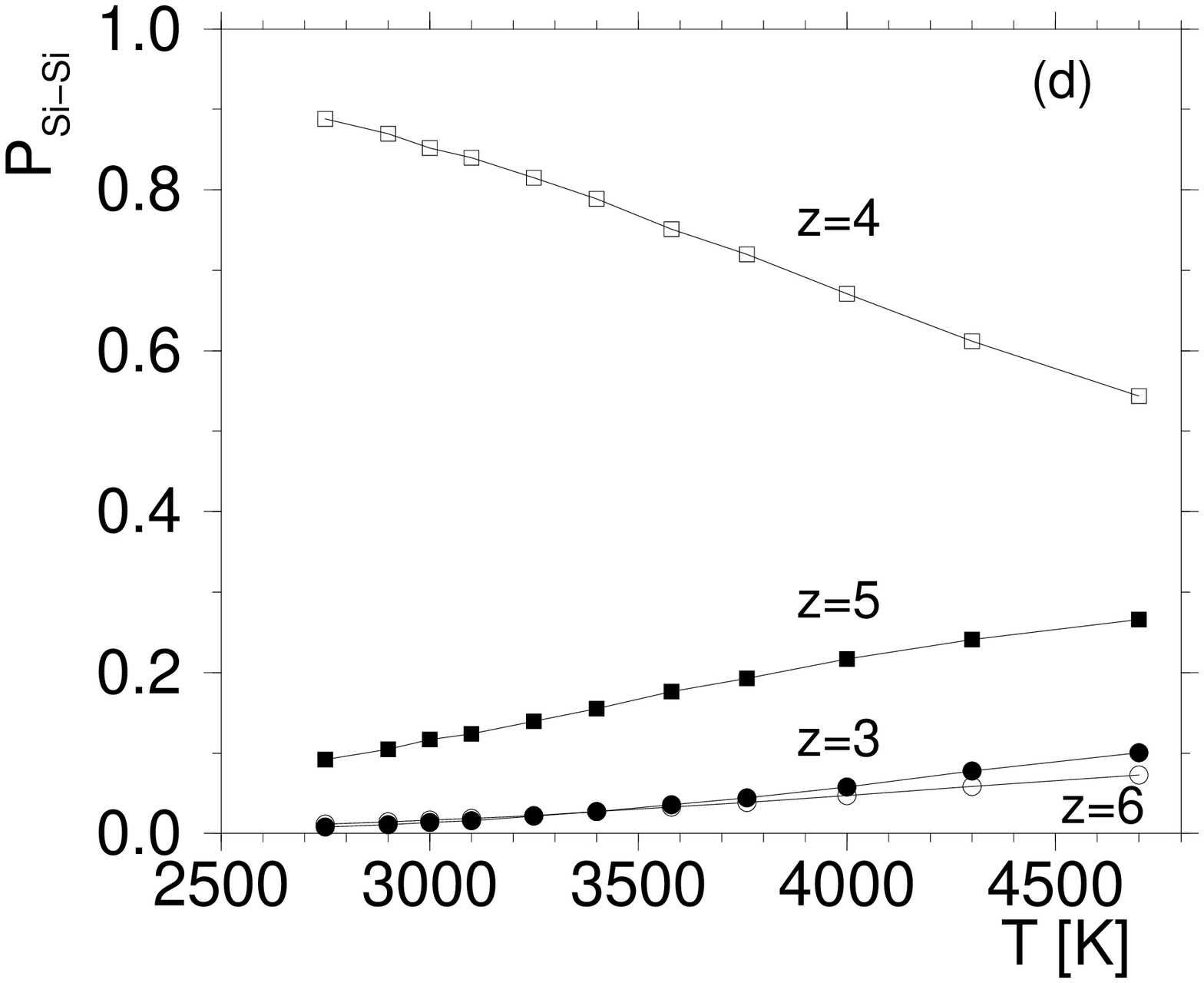,width=13cm,height=9.5cm}
\caption{Temperature dependence of the partial coordination numbers
$z$. The dashed lines in a) and b) are fits with an Arrhenius law. a)
$P_{\rm Si-O}$, b) $P_{\rm O-Si}$, c) $P_{\rm O-O}$, d) $P_{\rm
Si-Si}$.} \label{fig3}
\end{figure}

From the figure we also recognize that for all temperatures there is a
well defined minimum between the first and second peak. Therefore it is
possible to identify for each ion its nearest neighbors by requiring
that they are within the first nearest neighbor shell, defined by the
location of the mentioned minimum. For the following analysis we used
the values 3.64\AA, 2.35\AA, and 3.21\AA~for the location of this
minimum in the Si-Si-, Si-O-, and O-O-correlation.  In Fig.~\ref{fig3}
we show the probability $P_{\alpha\beta}$ that a ion of type $\alpha$
has exactly $z$ nearest neighbors of type $\beta$, for all relevant
values of $z$. From Figs.~\ref{fig3}a and \ref{fig3}b we recognize that
even at high temperatures more than 85\% of the silicon and oxygen
atoms are four and two fold coordinated, respectively, i.e. that they
are in the local environment expected from an ideal (corner-shared)
tetrahedral network.  At the lowest temperature investigated this
fraction is higher than 99\%. The remaining defects are silicon atoms
that are three and five fold coordinated and oxygen atoms that are
three and one fold coordinated, the latter thus forming dangling bonds.
We have found that for temperatures below 3700~K the probability for
the occurrence of such defects is described very well by an Arrhenius
law, $P_{\alpha\beta}=\pi_{\alpha\beta}\exp(-e_{\alpha\beta}/T)$
(dashed curves). As prefactors $\pi_{\alpha\beta}$ and activation
energies $e_{\alpha\beta}$ we found: $\pi_{\rm SiO}$=4.4, $e_{\rm
SiO}$=17130~K for $z=5$, $\pi_{\rm SiO}$=58.6, $e_{\rm SiO}$=31100~K
for $z=3$, $\pi_{\rm OSi}$=3.2, $e_{\rm OSi}$=17730~K for $z=3$ and
$\pi_{\rm OSi}$=8.9, $e_{\rm OSi}$=24760~K for $z=1$. From these
activation energies we recognize that at low temperatures defects in
which a silicon atom is coordinated by $z=5$ oxygen atoms or one in
which an oxygen atom is bound to $z=3$ silicon atoms are by far the
most frequent ones. In Refs.~\cite{vollmayr96b,vollmayr_diss} it was
shown that once the system has reached the glass transition temperature, the
distributions $P_{\rm SiO}$ and $P_{\rm OSi}$ do not change anymore,
i.e.,  that at low temperatures these distributions are given by the
ones at the glass transition temperature. By using the above given
activation energies we can therefore extrapolate the curves to the
experimental value of $T_g$ and thus estimate the density of the
defects of real silica below $T_g$.  Using the value
$T_g$=1450~K~\cite{bruckner70}, we predict that five fold coordinated
silicon atoms occur with probability $3.2\cdot 10^{-5}$ and three fold
coordinated oxygen atoms with probability $1.5\cdot 10^{-5}$.

Whereas the temperature dependence of the Si-O and O-Si defects is
quite simple, we see from Figs.~\ref{fig3}c and \ref{fig3}d that this
is not the case for the defects involving the structure on somewhat
larger length scales, since the functional form of the various curves is
not easy to identify. Thus the main information that can be obtained
from these figures is that locally the system seems to approach the
geometry of a random tetrahedral network. In this structure the most
probable configuration is that an oxygen atom has six other oxygen
atoms as its second nearest neighbors and a silicon atom has four other
silicon atoms as its second nearest neighbors.

\begin{figure}[h]
\psfig{file=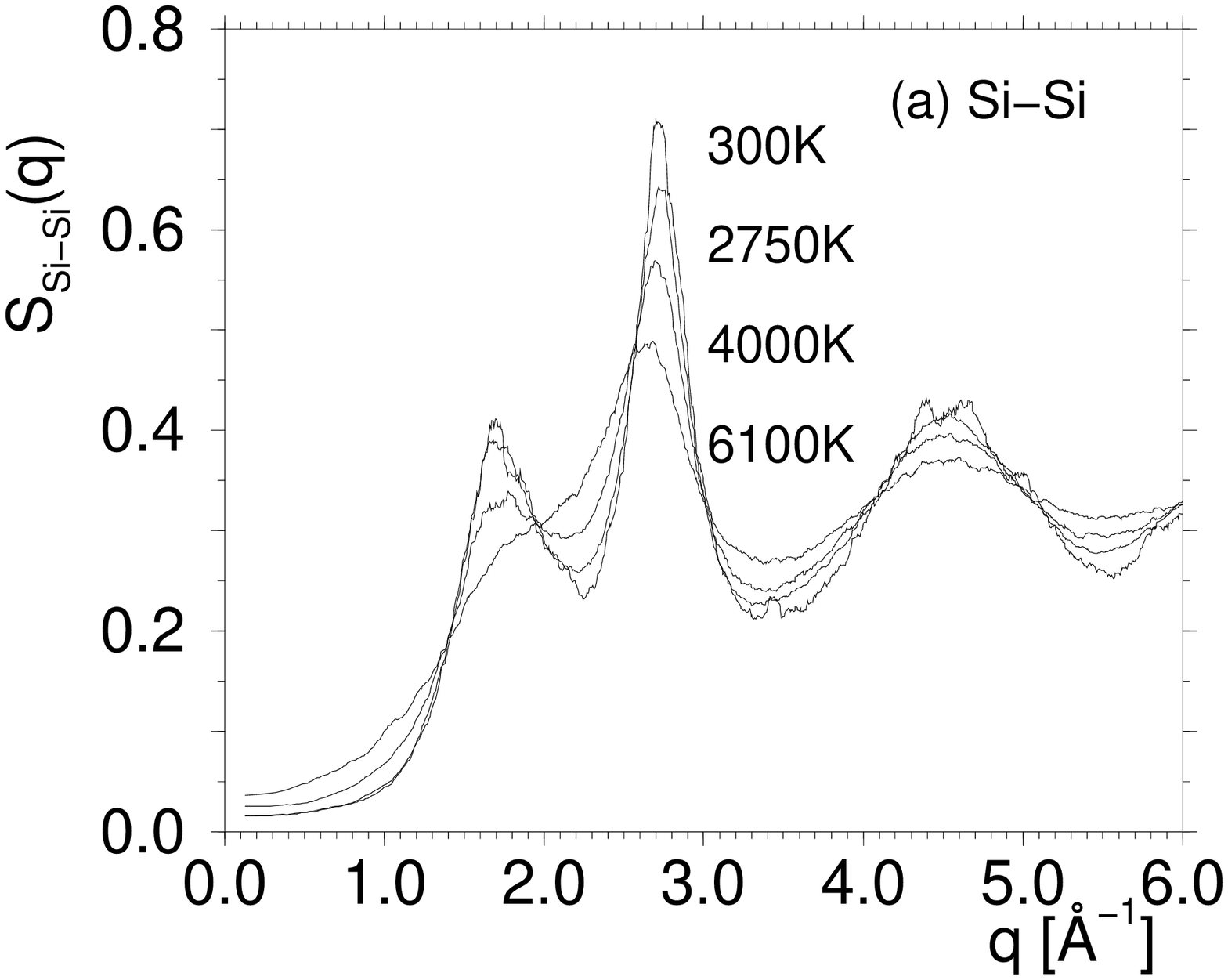,width=13cm,height=9.5cm}
\psfig{file=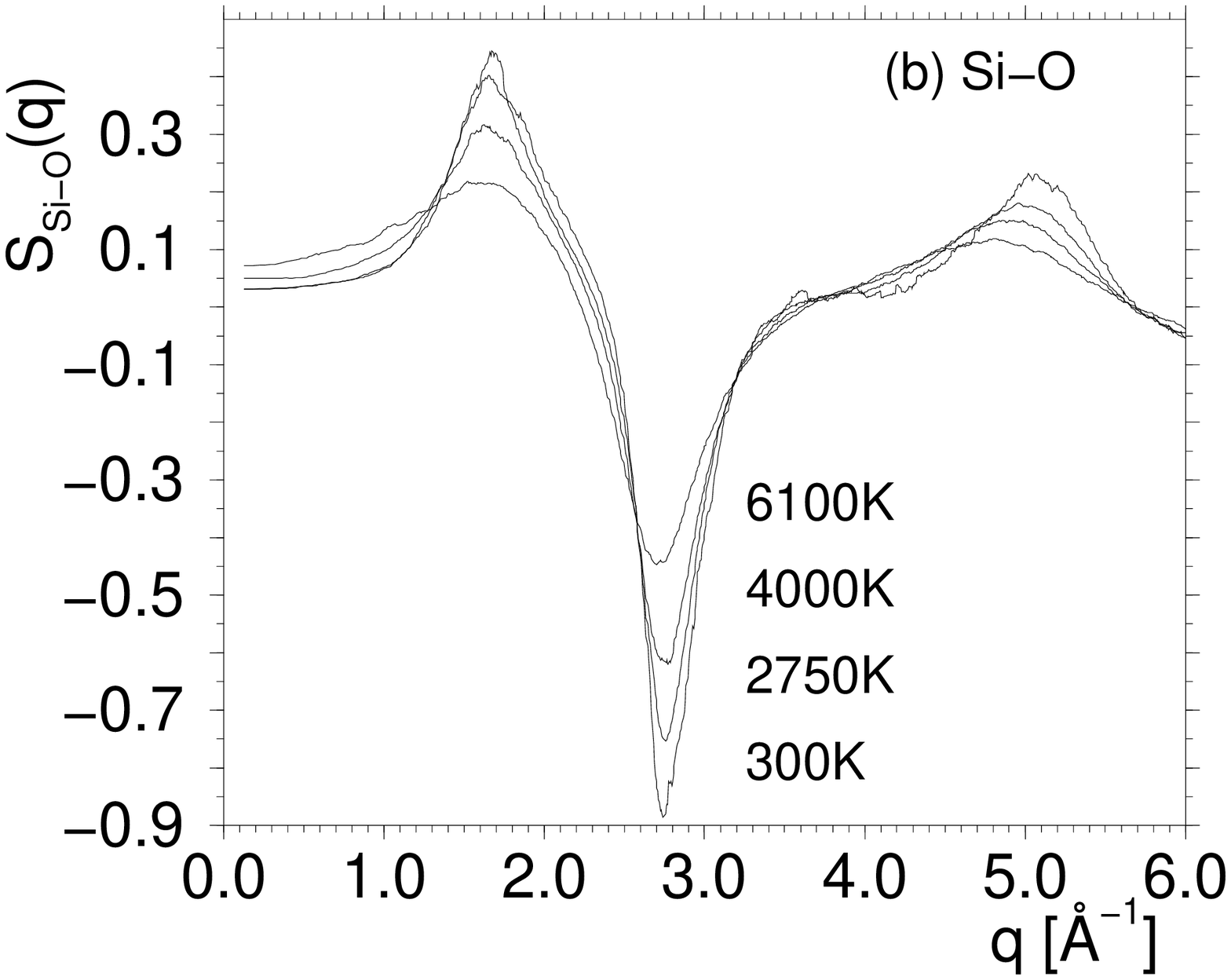,width=13cm,height=9.5cm}
\psfig{file=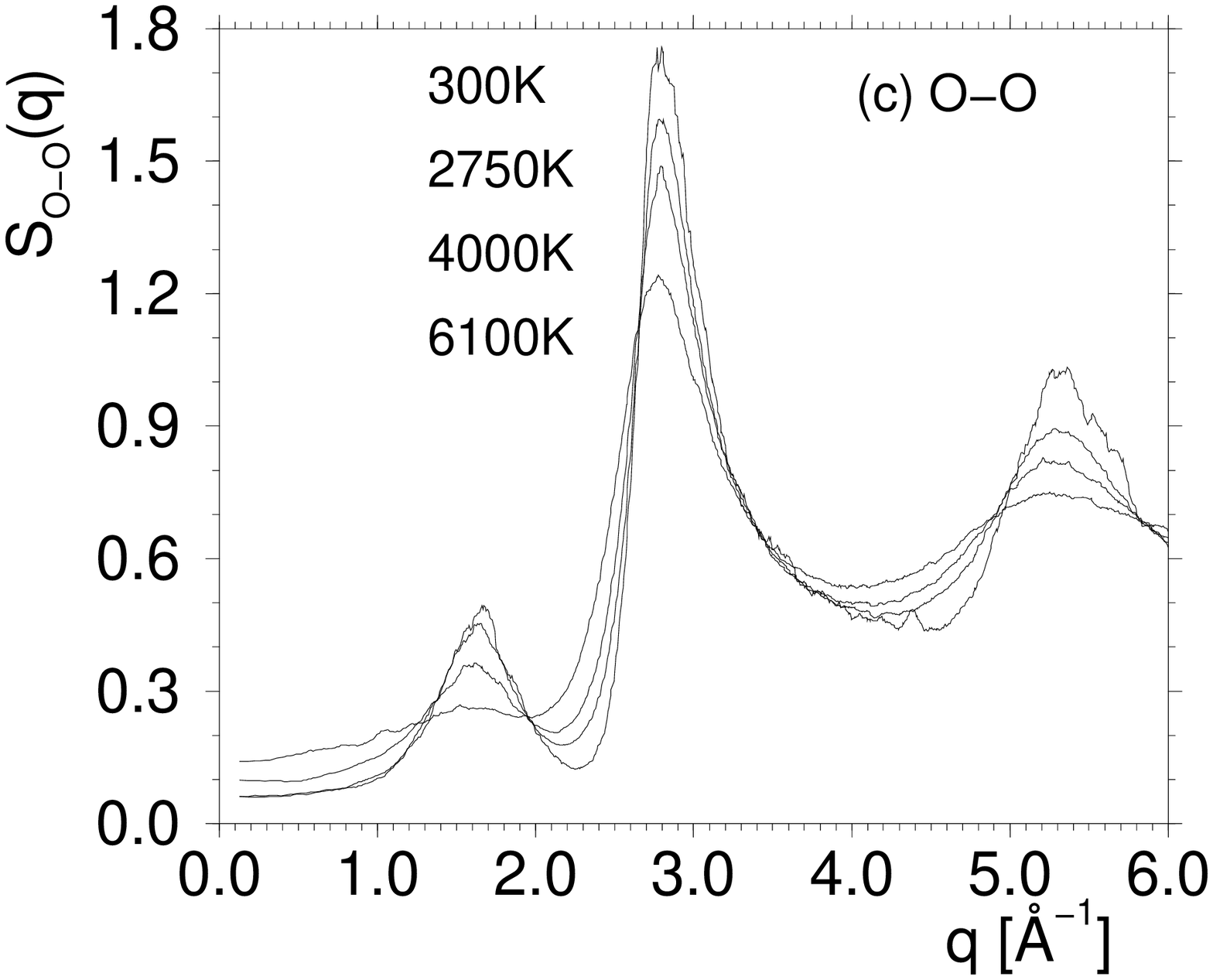,width=13cm,height=9.5cm}
\caption{Partial structure factors $S_{\alpha\beta}(q)$ for different
temperatures. The curves for 300~K were obtained by quenching the
system from a high temperature state and are therefore not equilibrium
curves. a) Si-Si, b) Si-O, c) O-O.}
\label{fig4}
\end{figure}

The structure at larger length scales can best be studied by means of
the partial structure factors $S_{\alpha\beta}(q)$~\cite{hansen86},
since any large scale feature will show up at wave-vectors $q$ which
are smaller than the location of the peak between neighboring
particles, i.e. in our case the Si-O correlation. In Fig.~\ref{fig4}
we show the three $S_{\alpha\beta}(q)$ for the same temperatures
already discussed in Fig.~\ref{fig2}. The peak corresponding to the
nearest neighbor Si-O distance is around 2.8\AA$^{-1}$. At smaller
wave-vectors an additional peak can be seen, the so-called first sharp
diffraction peak (FSDP). The microscopic reason for this peak is the
local chemical ordering of the ions in tetrahedra-like structures and
the location of the peak is related to the distance between neighboring
tetrahedra. The interesting information that can be obtained from
Fig.~\ref{fig4} is that the FSDP is observed already at temperatures as
high as 4000~K. At first glance it might seem a somewhat unrealistic
prediction of the BKS model that the open structure of the network
should be present even at such high temperatures. However, if one
recalls that even at temperatures as high as 2750~K silica is a quite
viscous liquid ($\eta \approx 2000$ Poise)~\cite{urbain82}, this result
is no longer that surprising, since such a high value of the viscosity
shows that even at these high temperatures the particles move quite
slowly, and that therefore their motion is strongly hindered by their
neighbors.  Therefore it is not surprising that the different
$S_{\alpha\beta}(q)$ show that there is indeed a lot of structure in
the system even at high temperatures.

Also included in Fig.~\ref{fig4} are curves for $T=300$~K. These were
obtained by using the equilibrated configurations at $T=2900$~K as
starting configurations of a cooling run in which the temperature was
decreased linearly within one million time steps to zero Kelvin. This
corresponds to a cooling rate of about 1.8$\cdot 10^{12}$~K/s.  With
this cooling rate the system falls out of equilibrium at around 2850~K
and thus this temperature corresponds to about the value of the fictive
temperature of the glass. The configurations that we obtained at
$T=300$~K were annealed for another 500,000 time steps before we
started to calculate the structure factors. From the figure we
recognize that the partial structure factors for this temperature seem
to evolve very smoothly from the (equilibrium) ones at higher
temperature, which however is not quite correct, as we will demonstrate
below.

\begin{figure}[h]
\psfig{file=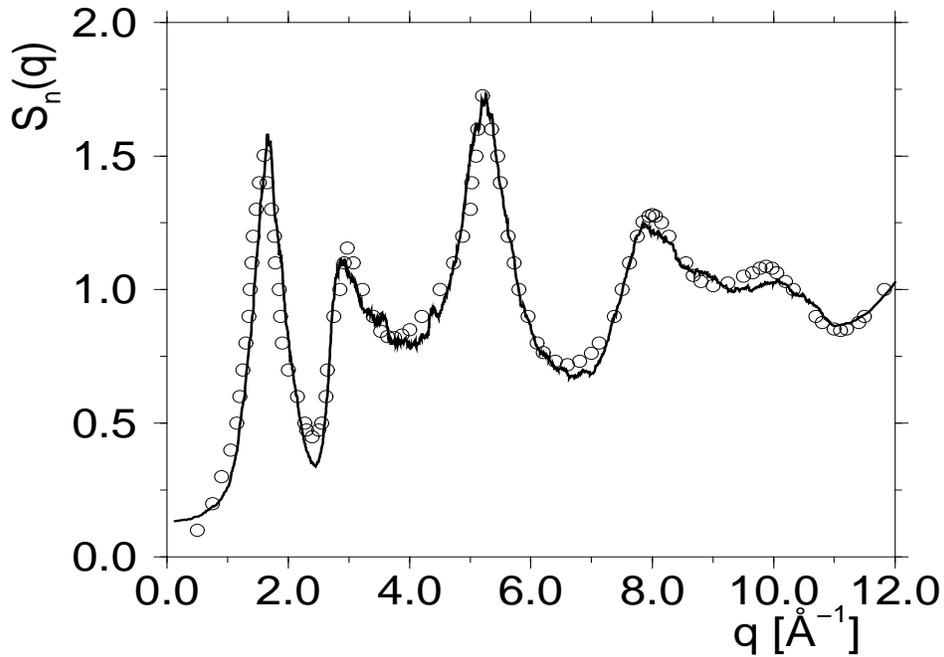,width=13cm,height=9.5cm}
\caption{Comparison of the neutron scattering function from our
simulation (solid line) with the experimental data of Price and
Carpenter~\protect\cite{price87} (circles).}
\label{fig5}
\end{figure}

\begin{figure}[h]
\psfig{file=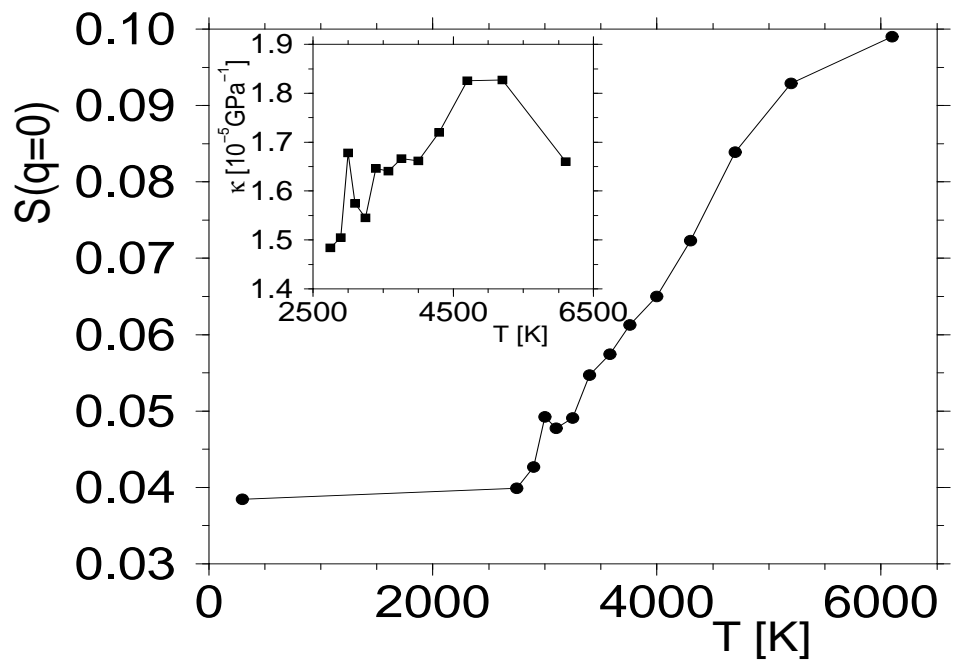,width=13cm,height=9.5cm}
\caption{Main figure: Value of the total structure factor at $q=0$ as a
function of temperature. Inset: temperature dependence of the
compressibility.}
\label{fig6}
\end{figure}

In order to compare our results for the structure factors with the one of
real silica we have also calculated the neutron scattering function
$S_n(q)$ from
\begin{equation}
S_n(q)=\frac{1}{N_{\rm Si}b_{\rm Si}^2+N_{\rm O}b_{\rm O}^2}
\sum_{k,l}^N b_k b_l \langle \exp(i{\bf q}\cdot 
({\bf r}_k -{\bf r}_l)) \rangle,
\label{eq2}
\end{equation}
where $b_k, k \in \{\rm Si,O\}$ are the neutron scattering cross
sections, and $\langle.\rangle$ is the thermal average.  Susman {\it et
al.}~\cite{susman91} report for $b_{Si}$ and $b_O$ the values
$0.4149\cdot 10^{-12}$~cm and $0.5803\cdot 10^{-12}$~cm, respectively.
Using these values and Eq.~(\ref{eq2}) we obtain for $T=300$~K the
$S_n(q)$ shown in Fig.~\ref{fig5}. Also included in the figure is the
neutron scattering result of Price and Carpenter at the same
temperature~\cite{price87}. We recognize that qualitatively, as well as
quantitatively, the two curves agree very well, in that not only the
location, but also the height of the various peaks is reproduced well.
This shows that the BKS potential is indeed able to reproduce this
structural quantity. A similar good agreement was obtained by Taraskin
and Elliott who calculated $S_n(q)$ for SiO$_2$ within the harmonic
approximation~\cite{taraskin97}. One significant difference between the
theoretical and the experimental curve can, however, be noticed, and
this is their behavior at small wave-vectors. Whereas an extrapolation
of the theoretical curve to $q=0$ seems to give a value around 0.16,
the experimental curve seems to extrapolate to a significantly smaller
value. It is well known that this value of the {\it total} structure
factor,
\begin{equation}
S(q)=N^{-1}\sum_{k,l}^N \langle \exp((i{\bf q}\cdot 
({\bf r}_k -{\bf r}_l)) \rangle,
\label{eq3}
\end{equation}
is related to the isothermal compressibility $\kappa_T$ of the system
via
\begin{equation}
\lim_{q\to 0} S(q)=\rho k_B T \kappa_T \quad,
\label{eq4}
\end{equation}
where $\rho=N/V$ is the particle density. We therefore extrapolated our
partial structure factors to $q=0$ and thus obtained $S(q)$ for $q\to
0$. This quantity is shown in Fig.~\ref{fig6} as a function of
temperature. We see that as long as we are in equilibrium, $T\geq
2750$~K, $S(q=0)$ decreases continuously with decreasing temperature.
At our (computer) glass transition $T\approx 2850$~K, however, the
structure freezes in and thus at $T=300$~K $S(q=0)$ is essentially
given by its value at this glass transition temperature\footnote{We
remind the reader that the glass at 300~K was produced by cooling the
system from 2900~K.}. Thus we see that it is not surprising that in
Fig.~\ref{fig5} the theoretical curve at small wave-vectors is above
the experimental one.

In the inset of Fig.~\ref{fig6} we show the compressibility of our
system as calculated from Eq.~(\ref{eq4}). We see that also this
quantity has a maximum at a temperature which corresponds to the
location of the density anomaly of the system, i.e.~for this model at
around 4700~K. No experimental data for $\kappa_T$ in this temperature
range are known to us. However, Fraser~\cite{fraser68} reports for
$T=1673$~K a compressibility around $2.7\cdot 10^{-5}$~GPa$^{-1}$.  If
we extrapolate the (equilibrium) compressibility data in
Fig.~\ref{fig6} to $T=1673$~K, we obtain a $\kappa_T$ at around
$1.35\cdot 10^{-5}$~GPa$^{-1}$ which is a factor of two smaller than
the one given by Fraser. Thus we have evidence that the BKS model is a
bit too stiff. The possibility that this discrepancy is due to the
slightly enhanced pressure of our system should, however, not be
disregarded.

\subsection{Dynamic properties}
\label{sec3.2}
Having studied the static properties of the silica melt we will
discuss in this subsection the dynamic ones.

One of the simplest quantities to study the dynamics of a fluid system
on a microscopic level is the mean squared displacement (MSD) $\langle
r^2(t) \rangle$ of a tagged particle (of type $\alpha$), which is given
by

\begin{equation}
\langle r^2(t) \rangle =\frac{1}{N_{\alpha}} \sum_{l=1}^{N_{\alpha}}
\langle |{\bf r}_l(t)-{\bf r}_l(0)|^2\rangle\quad .
\label{eq5}
\end{equation}

\begin{figure}[h]
\psfig{file=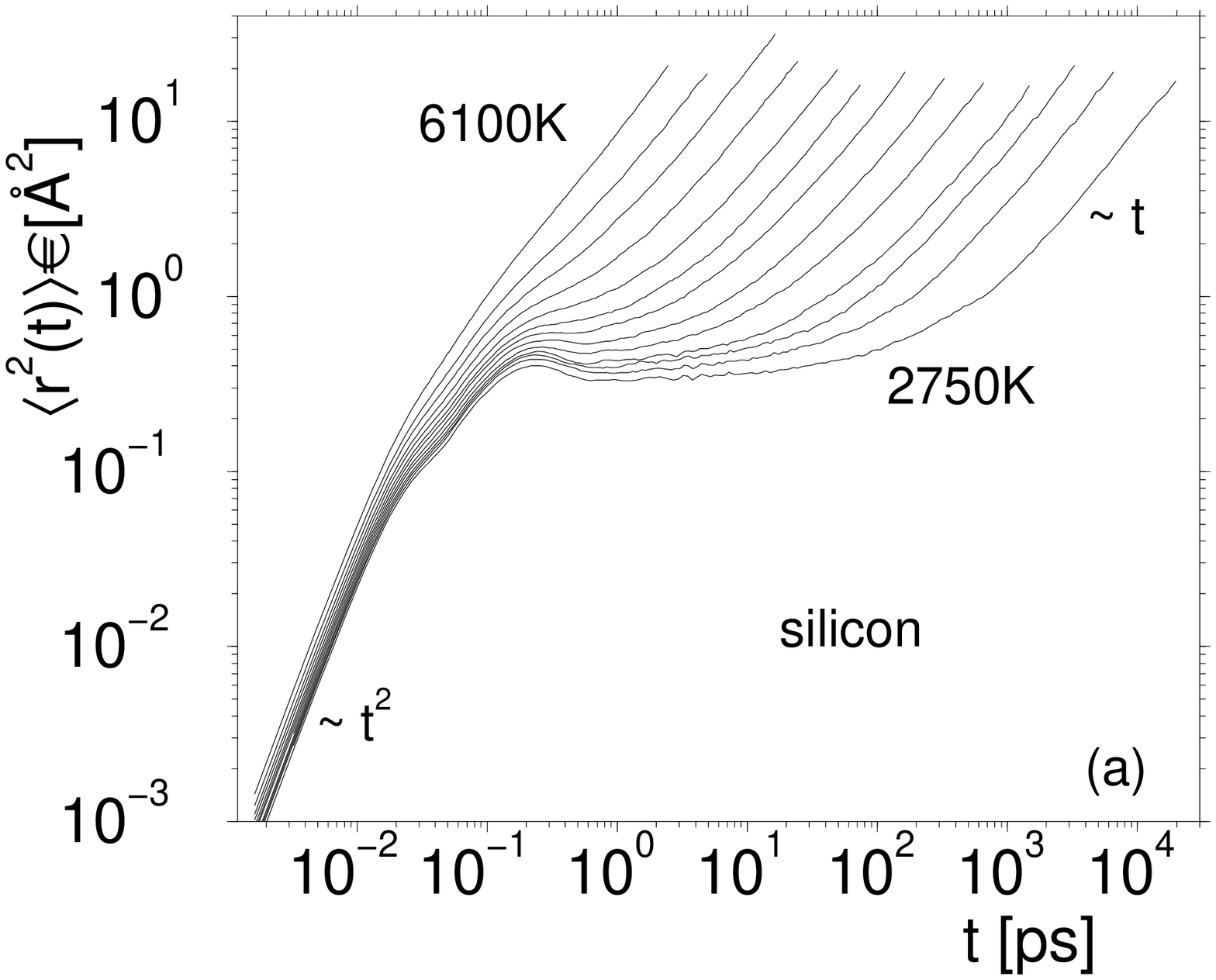,width=13cm,height=9.5cm}
\psfig{file=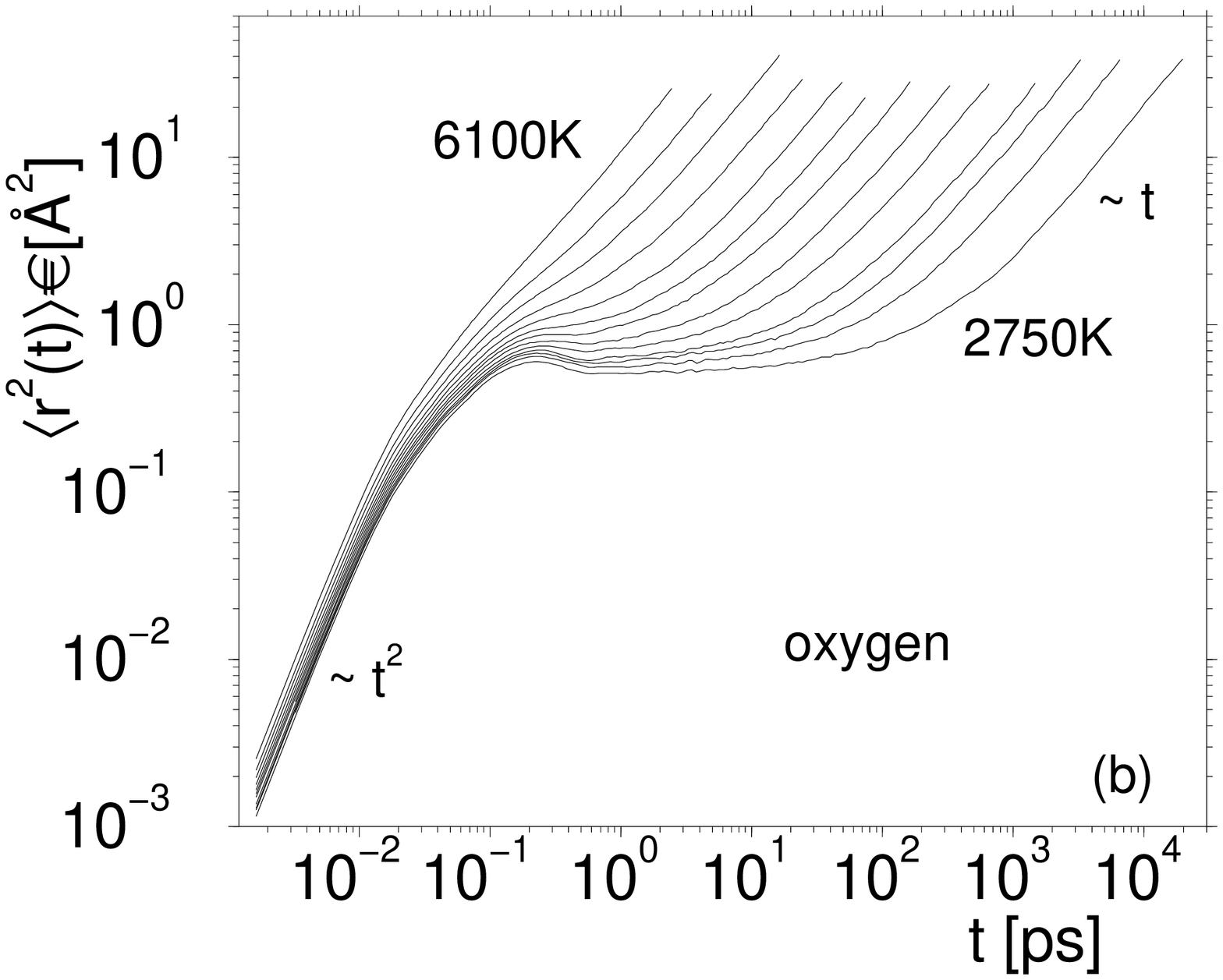,width=13cm,height=9.5cm}
\caption{Time dependence of the mean squared displacement for different
temperatures. a) silicon, b) oxygen.}
\label{fig7}
\end{figure}

The time dependence for this quantity is shown in Fig.~\ref{fig7} for
all temperatures investigated. For high temperatures, curves to the
left, two time regimes can be distinguished. At short times the motion
of the particles is ballistic, i.e.~${\bf r}_l(t)\approx {\bf r}_l(0)+
\dot{\bf r}_l t$, and the MSD is proportional to $t^2$. For long times the
motion is diffusive and hence we have $\langle r^2(t) \rangle \propto
t$. These two regimes are also seen at low temperatures, curves to the
right. However, at these temperatures these two regimes are separated
by a third one, in which the MSD does not change significantly over a
time span which extends over several decades in time. The microscopic
reason for this behavior is the so called cage-effect, i.e. the fact
that in this time regime the tagged particle is temporarily trapped by the
particles surrounding it and hence make an escape from its
cage very unlikely. Only for sufficiently large time the particle
succeeds to escape this (dynamic) cage and the MSD starts again to
increase significantly. The dynamics of the particles related to this
cage is usually called the $\beta$-relaxation and the mentioned
mode-coupling theory makes detailed predictions about this dynamics,
as it will be discussed in more details below.

What is also noticeable in the curves at low temperatures is that the
transition from the ballistic to the cage regime is quite different
from the one found in a simple liquid, see e.g. Ref.~\cite{kob95b}, in
that i) immediately after the ballistic regime the curves show a small
shoulder at around 0.03~ps and ii) a peak is observed at around
0.2~ps.  The first of these features, which is more pronounced in the
curves for the silicon atoms, is most likely the result of the complex
{\it local} motion of the atoms in the open tetrahedral network, such
as the bending and stretching of the tetrahedra. The second one is, as
already pointed out by Angell {\it et al.}~\cite{angell94}, related to
the so-called boson peak, a vibrational feature at low frequencies
whose precise origin is currently still a matter of
debate~\cite{taraskin97,horbach98b,bp_papers}.

We also point out that the above mentioned cage-effect becomes
observable already at temperatures as high as 3580~K. Usually this
effect is associated with the glass-former being in a {\it supercooled}
state. However, since the melting temperature $T_m$ of silica is around
2000~K, all our simulations are above $T_m$ and hence we do not
investigate the system in its supercooled regime at all. This shows
that the melting temperature is, from the point of view of the cage
effect, an irrelevant temperature. That this somewhat surprising result
is not a particularity of the silica model studied here can be inferred
from the fact that experimental studies of glycerol and B$_2$O$_3$ also
show a strong non-Debye relaxation at temperatures well above their
melting temperature~\cite{glycerol,rossler94}. All these results give
support to the point of view of the earlier mentioned MCT, that the
whole slowing down of the system is a purely kinetic phenomenon and
thus not related to any thermodynamic singularity of any kind.

\begin{figure}[h]
\psfig{file=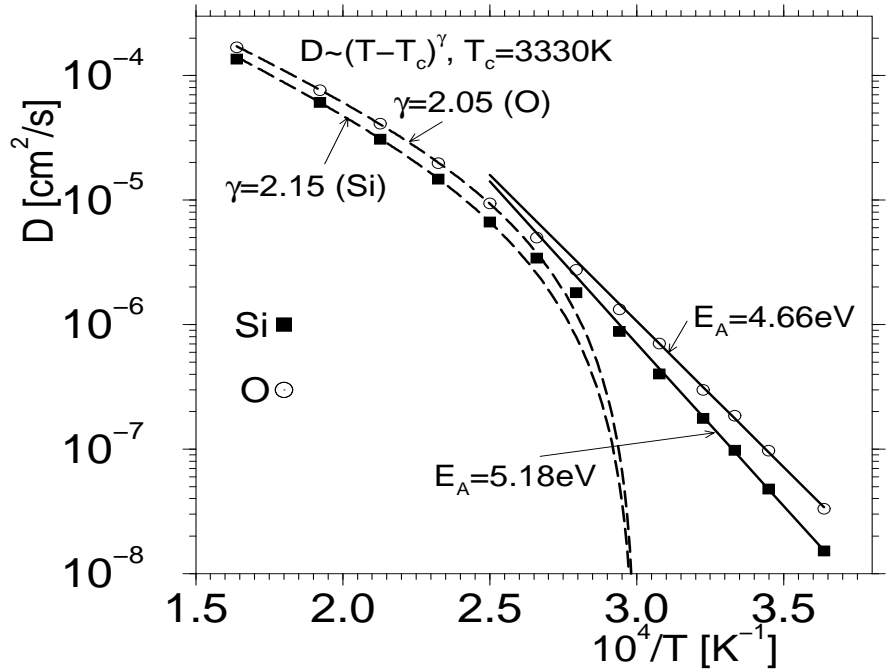,width=13cm,height=9.5cm}
\caption{Arrhenius plot of the diffusion constants. Bold solid lines:
Arrhenius fits to the data at low temperatures with the stated
activation energies. Dashed lines: Results of a fit to the high
temperature data with the power-law predicted by MCT with a critical
temperature of 3330~K.}
\label{fig8}
\end{figure}

Using the Einstein relation $\lim_{t\to \infty} \langle r^2(t)/6t
\rangle=D$, it is easy to calculate the diffusion constants $D$ from
the MSD. These are plotted in Fig.~\ref{fig8} as a function of the
inverse temperature. As expected for a strong glass-former, we find
that at low temperatures the diffusion constants show an Arrhenius
dependence. The activation energies are 4.66~eV and 5.18~eV for oxygen
and silicon, respectively (bold solid lines). These numbers compare very
well with the ones determined in experiments at significantly lower
temperatures, namely 4.7~eV for oxygen~\cite{mikkelsen84} and 6~eV for
silicon~\cite{brebec80}. That such an excellent agreement between the
theoretical and experimental activation energies is not trivial has
recently been demonstrated by Hemmati and Angell~\cite{hemmati98}, who
showed that various models for silica can give rise to quite different
activation energies. These authors also showed that the different
models predict diffusion constants which differ by up to two decades at
temperatures as high as 3000~K, which shows that dynamical quantities
like $D$ depend much more sensitively on the potential than 
structural quantities.

For higher temperatures we see from the figure that significant
deviations from the Arrhenius behavior are observed in that the
diffusion constants increase slower with increasing temperature
than expected from an activated process. A similar change in the
temperature dependence of $D$ has also been found in many other
models for silica~\cite{hemmati98}. Recently Hess {\it et al.} have
reported the analysis of experimental viscosity data of SiO$_2$ at high
temperatures and find that deviations from a pure Arrhenius law are
present~\cite{hess96}.

One explanation to rationalize the observed deviations from the
Arrhenius behavior is offered by MCT~\cite{mct}, since the so-called
ideal version of this theory predicts that the temperature dependence
of the diffusion constant, as well as the inverse of the
$\alpha-$relaxation time $\tau$(T), is given by a power-law, i.e.
\begin{equation}
D\propto \tau^{-1} \propto (T-T_c)^{\gamma} \quad ,
\label{eq6}
\end{equation}
where the critical temperature $T_c$ and the critical exponent $\gamma$
can, in principle, be calculated from the temperature dependence of the
partial structure factors.  In practice, however, the two quantities
are usually taken as fit parameters~\cite{mct} (for exceptions see
\cite{nauroth97} and references therein).  Instead of using the
diffusion data alone to determine $\gamma$ and $T_c$, it has been found
in similar types of analysis (e.g.
\cite{kob_lj,kob95b,sciortino,kammerer}) that more reliable results are
obtained if one determines these quantities from simultaneous fits to
the diffusion constants as well as to the $\alpha-$relaxation times for
different wave-vectors, and therefore we proceeded in this way
too~\cite{horbach99}. The result of these fits is that the critical
temperature is 3330~K. This value of $T_c$ is in excellent agreement
with the one determined by Hess {\it et al.} in their analysis of
viscosity data of real silica, which is 3221~K~\cite{hess96}. (We note
that we have learned about the results of Hess {\it et al.} only after
having determined our value of $T_c$). Using our value of $T_c$, we
obtain for the critical exponent $\gamma$ 2.15 and 2.05 for the silicon
and oxygen diffusion constant, respectively. Note that the theory
predicts that the value of $\gamma$ should be independent of the
species and the fact that the two values we find are so close together
supports this prediction.  The so obtained power-law fits are included
in Fig.~\ref{fig8} as well and we see that they give a good description
of the diffusion constants over about 1.5 decades in $D$. This range is
significantly smaller than the one found in the case of simple liquids
for which the power-law can be observed over about three
decades~\cite{roux89,kob_lj}. As substantiated below, the reason for
this relatively small range is very likely the fact that in SiO$_2$ the
relaxation dynamics at low temperatures is dominated by strong hopping
processes, i.e.~jump like motions of the ions. Since the {\it ideal}
version of MCT, i.e. the version that predicts the power-law given by
Eq.~(\ref{eq6}), does not take into account these types of processes,
it is thus not surprising that this version of the theory is applicable
only in a quite limited temperature range. The solution to this problem
might be the so-called extended MCT, i.e. that version of the theory in
which such hopping processes are taken into
account~\cite{mct,hopping_mct}. Since, however, so far the details of
this theory for the $\alpha$-relaxation have not been worked out, no
test can presently be made. This situation is different in the
$\beta-$relaxation regime, where some of the predictions of the
extended MCT have been worked out and as we will demonstrate in a
different place~\cite{horbach99}, that these predictions hold very well for
the present system.

Although Fig.~\ref{fig7} suggests that MCT might be useful to describe
the dynamics of the present system, the power-law fits in
Fig.~\ref{fig8} are of course not a proof that the extracted parameters
$T_c$ and $\gamma$ have a deep physical meaning. The significance of
the theory for SiO$_2$ becomes only obvious if also the $\alpha$- and
$\beta$-relaxation dynamics of the time correlation functions are
analysed carefully, which is done in
Refs.~\cite{horbach_diss,horbach99}.  We mention already at this place,
however, that the value for the critical exponent found for the
diffusion constant, $\gamma\approx 2.1$ is slightly smaller than the
one found for the $\alpha$-relaxation time $\tau$ for larger
wave-vectors, which is around 2.35~\cite{horbach99}. The value of this
latter exponent is compatible with the relation proposed by MCT between
these critical exponents and the value of the so-called von Schweidler
exponent of the $\beta$-relaxation regime. Hence we conclude that the
critical exponent of the diffusion constant is smaller than expected
from MCT, an observation which is in agreement with the one made for
simple liquids~\cite{kob_lj,gleim99}.

\begin{figure}[h]
\psfig{file=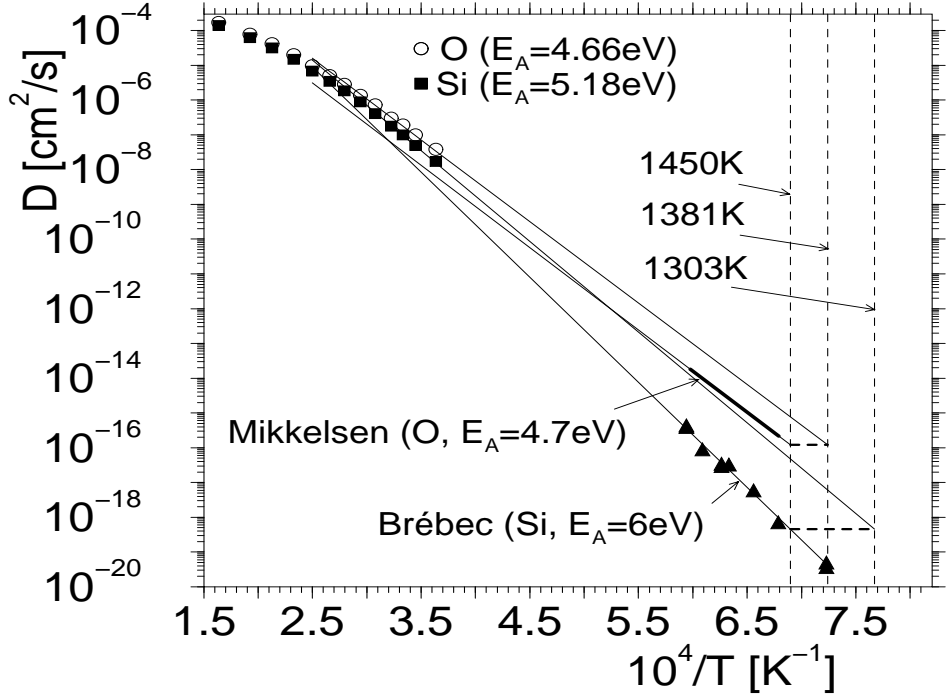,width=13cm,height=9.5cm}
\caption{Arrhenius plot of the diffusion constants as determined from
our simulation (open circles, oxygen; filled squares, silicon) and the
experimental values for oxygen (bold line)~\protect\cite{mikkelsen84}
and silicon\protect\cite{brebec80}. Thin solid lines: Extrapolation
of our data to low temperatures and extrapolation of the experimental
data to high temperatures. See text for the discussion of the
other lines.}
\label{fig9}
\end{figure}

Having now presented the temperature dependence of the diffusion
constants in the temperature range of our simulation, it is instructive
to compare these results with the one of real experiments. For this we
show in Fig.~\ref{fig9} the diffusion constants as measured by us and
the ones determined in experiments by Mikkelsen~\cite{mikkelsen84} and
by Br\'ebec {\it et al.}~\cite{brebec80}. Also included are the
Arrhenius fits to our data. We see that the extrapolation of these fits
to the temperature range which is accessible to the experiments
overestimates the diffusion constant by about one decade in the case of
oxygen and by about two decades in the case of silicon.
From such a plot we can also estimate the temperature at which the BKS
model would show the {\it experimental} glass transition temperature
$T_g$, if we would have access to computers which are by about a factor
of $10^{10}$ faster.  For this we read off the experimental values of
the diffusion constants at the experimental $T_g$=1450~K (horizontal
dashed lines) and determine the temperatures at which the Arrhenius
extrapolation of our data crosses these values. From this construction
we obtain a $T_{g,sim}$ of 1380~K and 1303~K for oxygen and silicon,
respectively (vertical dashed lines). Thus we come to the remarkable
conclusion that using the BKS potential it is possible to estimate the
experimental glass transition temperature to within 10\%.

Another important transport quantity is the shear viscosity $\eta$.
Since $\eta$ is a collective quantity, it is quite demanding to measure
it in a simulation with high accuracy. For the case of silica the only
simulation we know in which $\eta$ has been determined is the one by
Barrat {\it et al.} in which the pressure and temperature dependence of
the viscosity was determined at relatively high
temperatures~\cite{barrat97}. 

As in the case of the diffusion constant there are two possibilities
to calculate $\eta$: from a Green-Kubo relation and
a generalized Einstein relation~\cite{boon_yip80}. The Green-Kubo
relation is

\begin{equation}
\eta=\frac{1}{k_BTV}\int_0^{\infty} dt \langle
\dot{A}_{\alpha\beta}(t) \dot{A}_{\alpha\beta}(0) \rangle \quad,
\label{eq7}
\end{equation}
where the off-diagonal elements of the pressure tensor are given by
\begin{equation}
\dot{A}_{\alpha\beta}=\sum_{i=1}^{N}m_i v_i^{\alpha} v_i^{\beta} +
\sum_{i=1}^N \sum_{j>i}^N F_{ij}^{\alpha} r_{ij}^{\beta}\quad
\alpha \neq \beta.
\label{eq8}
\end{equation}
Here $F_{ij}^{\alpha}$ is the $\alpha$ component of the force between
ions $i$ and $j$.

The Einstein-formula reads:
\begin{equation}
\eta=\frac{1}{k_BTV}\lim_{t\to \infty} \langle
(A_{\alpha\beta}(t)-A_{\alpha\beta}(0))^2 \rangle \quad
\label{eq9}
\end{equation}
with
\begin{equation}
A_{\alpha\beta}(t)=\sum_{i=1}^N m_i v_i^{\alpha}(t) r_i^{\beta}(t)\quad .
\label{eq10}
\end{equation}

Allen {\it et al.}~have pointed out that in order to obtain correct
results it is important to represent the term
$A_{\alpha\beta}(t)-A_{\alpha\beta}(0)$ in Eq.~(\ref{eq9}) in a way
which is independent of the coordinate system~\cite{allen94}.  One
possible choice is

\begin{equation}
A_{\alpha\beta}(t)-A_{\alpha\beta}(0)=\int_o^t dt'
\dot{A}_{\alpha\beta}(t').
\label{eq11}
\end{equation}

We also mention that care has also been taken in the evaluation of
the forces $F_{ij}^{\alpha}$ occurring in Eq.~(\ref{eq8}), since it is
not possible to use just the ones obtained from the Ewald sums. More
details on this problem can be found in Refs.~\cite{visc_ewald}. In the
course of our calculation we found that the two methods discussed to 
calculate $\eta$ give results with similar accuracy, and thus can be
considered to be equivalent from a numerical point of view.

Because of the mentioned collective nature of $\eta$, the usual two
runs we did at every temperature were not sufficient to allow to
determine $\eta$ with a satisfactory statistical accuracy.  Therefore
we performed for all temperatures above 2900~K 18 extra runs in order
to calculate $\eta$. Because of the long relaxation times at $T=2900$~K
and $T=2750$~K it was not possible to do that many runs at these
temperatures and thus no results for $\eta$ have been obtained.

\begin{figure}[h]
\psfig{file=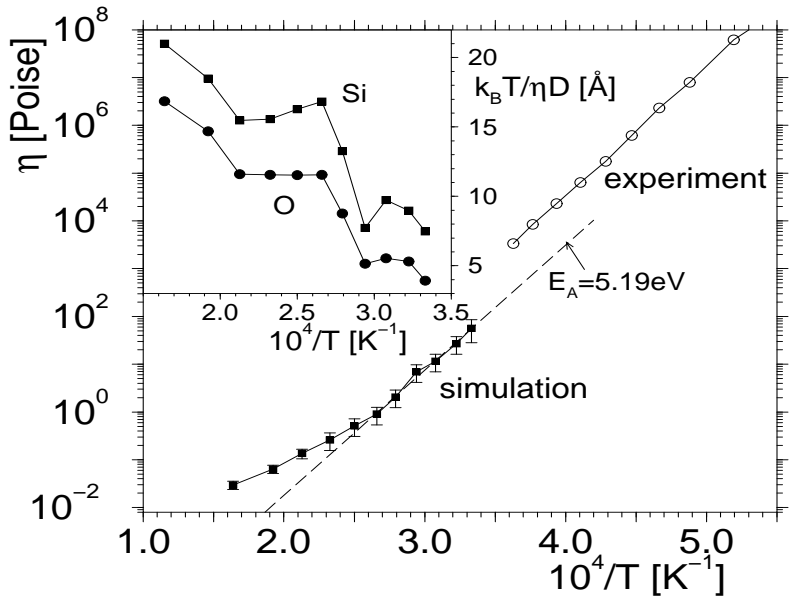,width=11cm,height=8.0cm}
\caption{Main figure: Arrhenius plot of the shear viscosity from the
simulation (filled squares). The dashed line is a fit with an Arrhenius
law to our low-temperature data. 
The open circles are experimental data from Urbain
{\it et al.}\protect\cite{urbain82}. Inset: Temperature dependence of
the left hand side of Eq.~(\protect\ref{eq12}) to check the validity of
the Stokes-Einstein relation.}
\label{fig10}
\end{figure}

In Fig.~\ref{fig10} we show the temperature dependence of the viscosity
in an Arrhenius plot (filled squares in main figure). We see that,
similar to our results for the diffusion constants, also this transport
quantity shows at low temperatures an Arrhenius behavior which crosses
over to a weaker temperature dependence with increasing temperature. In
the temperature regime in which the Arrhenius law is observed the data
is described very well by an Arrhenius law. If we fit the six lowest
temperatures with such a law we find an activation energy of 5.19~eV,
which is in very good agreement with the experimental value of
5.33~eV~\cite{urbain82}. By using such an Arrhenius law to extrapolate
our data to lower temperatures we see that for temperatures around
2800~K such an extrapolation underestimates the viscosity of real
silica, as measured by Urbain {\it et al.}~\cite{urbain82} (open
circles), by about a factor of 10. Thus we find that the BKS model
underestimates the viscosity of real silica but that the error lies in
the prefactor and not in the activation energy. (We also mention that
there seems to be an uncertainty of the experimental value of the
prefactor of about a factor of two~\cite{mazurin83}, thus it might be
that the actual discrepancy between our simulation and reality is less
than just stated.) We also mention that if the Arrhenius law found for
our low temperature data is used for estimating the temperature at
which the viscosity of our system reaches the value $10^{13}$~Poise,
i.e. the value of the experimental glass transition temperature, we
find that this value is at 1310~K. Thus we find, in agreement with our
results from the diffusion constant, that the BKS model underestimates
the experimental glass transition temperature by about 10\%.

Having an independent measurement of the diffusion constants and
the viscosity, it is possible to check the validity of the
Stokes-Einstein relation in this system. This relation is given by
\begin{equation}
\frac{k_BT}{\eta D}=\lambda=const \quad,
\label{eq12}
\end{equation}
where the constant $\lambda$ has, in the phenomenological Eyring
theory, the meaning of a length of an elementary diffusion
step~\cite{eyring41}. To check whether the left hand side of
Eq.~(\ref{eq12}) is indeed constant we plot its temperature dependence
in the inset of Fig.~\ref{fig10}. From this inset we see that $\lambda$
depends in the whole temperature interval on $T$ in that it changes
from values around 20\AA~at high temperatures to values around 5\AA~at
the lowest temperatures. This finding is not surprising in view of the
fact that the activation energy of the viscosity (5.19~eV, see
Fig.~\ref{fig10}) is very close to the one of the diffusion constant of
silicon (5.18~eV, see Fig.~\ref{fig8}), thus showing that for silicon
the product $\eta D_{\rm Si}$ is constant, whereas in the
Stokes-Einstein relation an additional factor $T$ is present.  Thus we
conclude that in the temperature range considered here, the
Stokes-Einstein relation is not a good way to convert viscosity data
into diffusivities or vice versa.

\begin{figure}[h]
\psfig{file=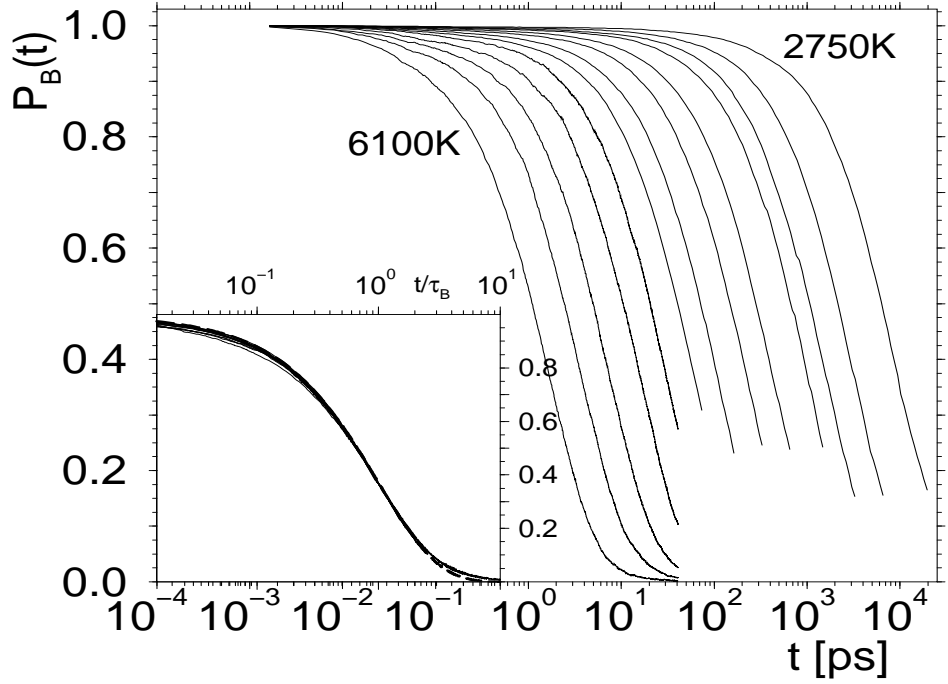,width=13cm,height=9.5cm}
\caption{Time dependence of $P_B$, the probability that a bond between
a silicon and an oxygen atom which exists at time zero is also present at
time $t$, for all temperatures investigated (main figure). Inset:
The same data versus the rescaled time $t/\tau_B$, where $\tau_B(T)$
is the decay time of $P_B$. The dashed line is a fit with a KWW-law
with $\beta=0.90$.}
\label{fig11}
\end{figure}

From our data on the temperature dependence of the diffusion
constants and the viscosity we have evidence that with increasing
temperature the relaxation behavior of the system crosses over from one
similar to an activated process to one which is, potentially, described
well by the ideal version of MCT. On the other hand, the ideal version
of the theory predicts a power-law divergence to infinity of $D$ and
$\eta$ at $T_c$, which in our case is at 3330~K, while Figs.~\ref{fig8}
and \ref{fig10} show that this is not the case. Thus the need for the
extended version of MCT is evident.  Further support for a change of
the transport mechanisms is obtained by investigating the temperature
dependence of the life time of a bond between a silicon and an oxygen
atom. From Fig.~\ref{fig2}b we see that even at the highest temperature
there is a well defined minimum between the first and second neighbor
peak in the radial distribution function $g(r)$ for the Si-O
correlation. Hence it is quite natural to make the definition that a
silicon and an oxygen atom are bonded if their separation is less than
the location of this minimum, which we located at 2.35\AA. In
Fig.~\ref{fig11} we show $P_B(t)$, the probability that a bond which
was present at time zero is still present at time $t$. From this figure
we see that the decay time of this probability increases fast with
decreasing temperature. More interesting is the observation that at low
temperatures $P_B(t)$ does not decay to zero within the time span of
our simulation, but that the curves end at around a value of 0.2. In
Fig.~\ref{fig7} we have shown that our simulations have been long
enough to allow to observe the diffusive regime in the mean squared
displacement and in Refs.~\cite{horbach98a,horbach99} we show that also
the structural correlation functions, such as the intermediate
scattering functions $F_s(q,t)$ and $F(q,t)$ at the
FSDP~\cite{boon_yip80}, decay to zero within the time span of our
simulation. Hence we conclude that in order to become diffusive, or for
the decay of the mentioned structural correlation functions, it is not
necessary that all of the bonds between silicon and oxygen
break\footnote{We mention that this is the case at high temperatures
also. However, since at these temperatures the relaxation times $\tau$
are short our runs were so long, compared to $\tau$, that in the time
span of the simulation even $P_B(t)$ decays to zero.}.

From the figure one also sees that the shape of the curves does not
seem to depend on temperature. This can be checked by plotting
$P_B(t)$ versus $t/\tau_B$, where the ``life time''$\tau_B(T)$ is
defined by requiring that $P_B(\tau_B)=e^{-1}$. In the inset of
Fig.~\ref{fig11} we show this type of plot and we see that the curves
for the different temperatures fall indeed very well on top of each
other. At short and intermediate times the master curve is approximated
well by a Kohlrausch-Williams-Watts curve (KWW),
$\exp(-(t/\tau_B)^{\beta})$, with $\beta=0.90$ (dashed line).  decay at
short times and a power-law decay at long times.

\begin{figure}[h]
\psfig{file=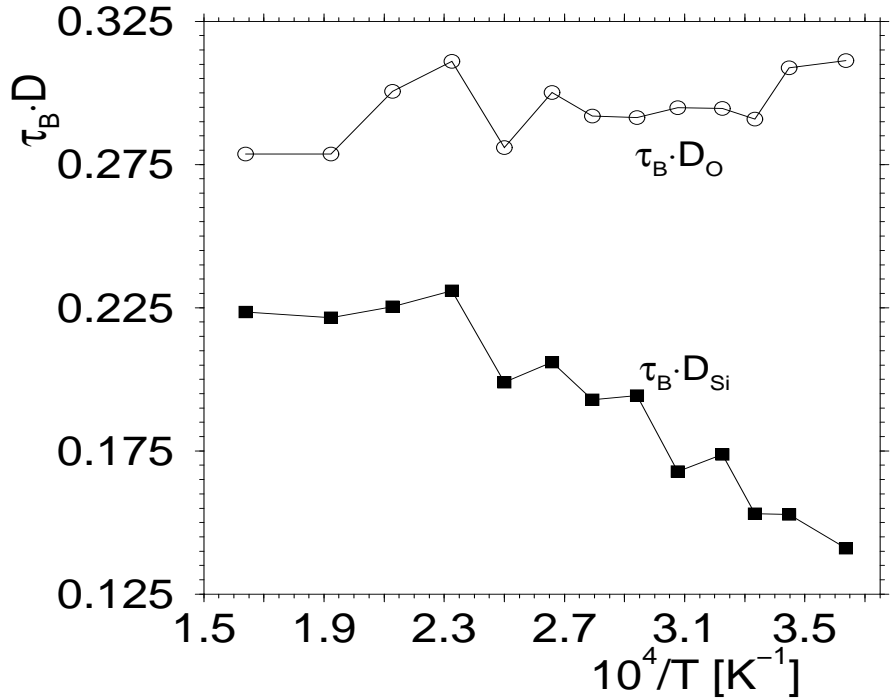,width=13cm,height=9.5cm}
\caption{Temperature dependence of the products of the diffusion
constants with the decay time of $P_B$.}
\label{fig12}
\end{figure}

In order to find out how the breaking of the bonds is related to the
diffusion process we plot in Fig.~\ref{fig12} the product of the life
time $\tau_B$ and the diffusion constants versus the inverse
temperature. From this figure we recognize that the product $\tau_B
D_O$ is essentially independent of temperature, demonstrating that the
breaking of the bond is indeed related to the elementary diffusion step
of the oxygen atoms. For the silicon atoms the situation is more
complicated in that the product $\tau_B D_{\rm Si}$ is constant at high
temperatures but then starts to decrease at low temperatures.  Thus we
have evidence that at high temperatures the diffusion mechanism for the
silicon atoms is very similar to the one of the oxygen atoms and is
likely governed by the collective motion described by MCT. When the
system enters the Arrhenius regime the situation changes in that now
the silicon atoms see a quite different local (mean) potential than the
oxygen atoms (since the former sit in deeper minima) and thus their
diffusion motion is not directly related to the breaking of the bond
with one of its neighbors.

\begin{figure}[h]
\psfig{file=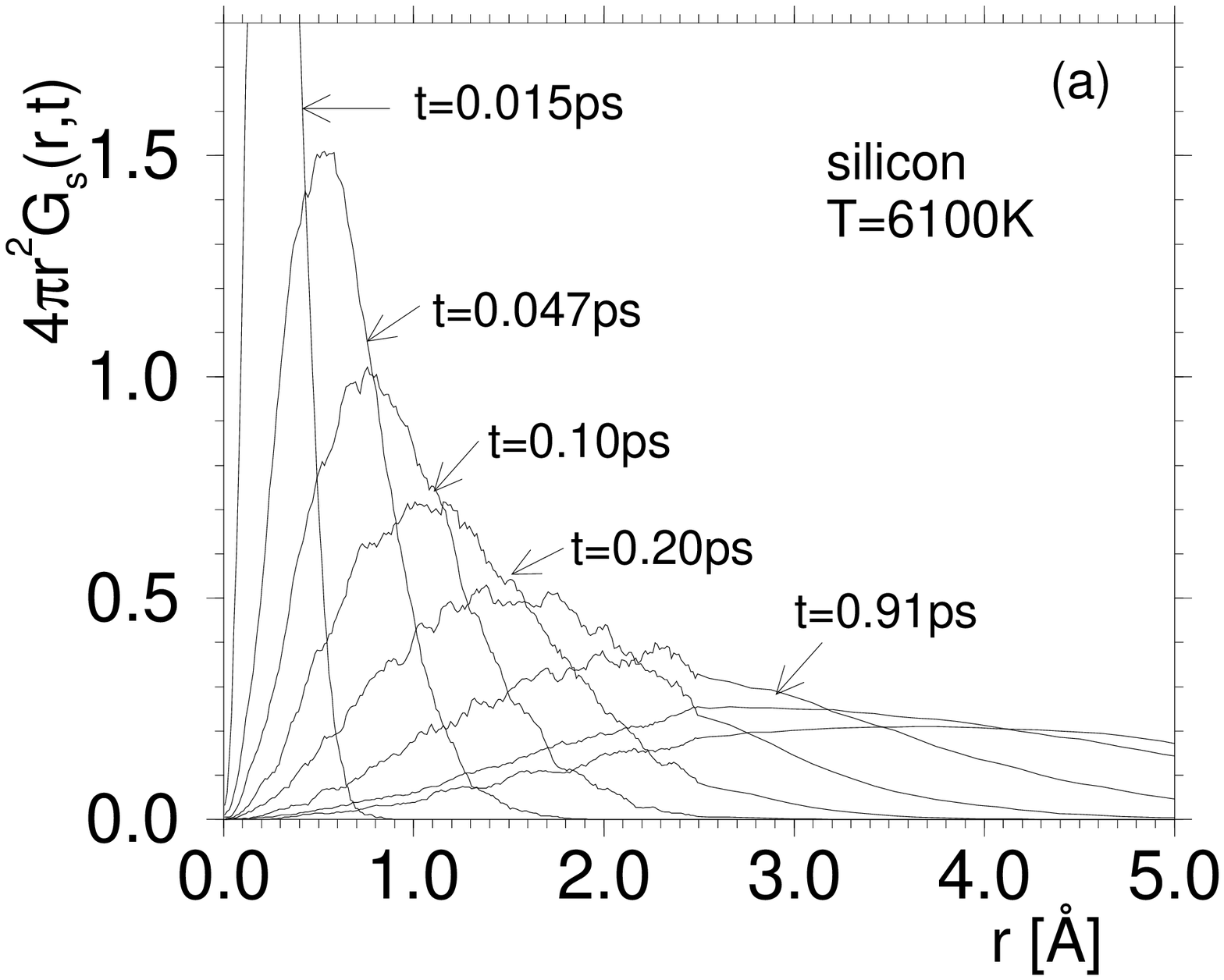,width=13cm,height=9.5cm}
\psfig{file=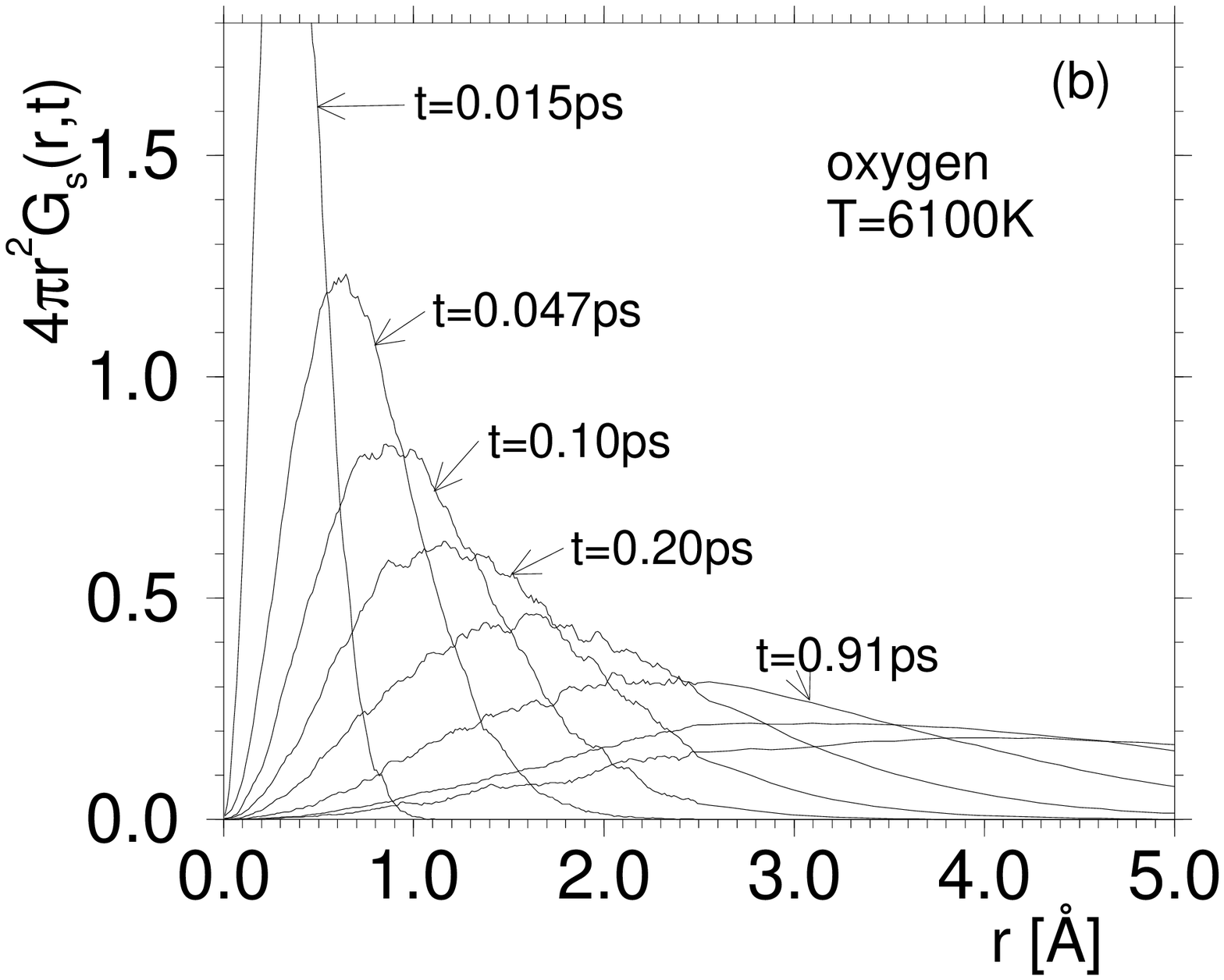,width=13cm,height=9.5cm}
\psfig{file=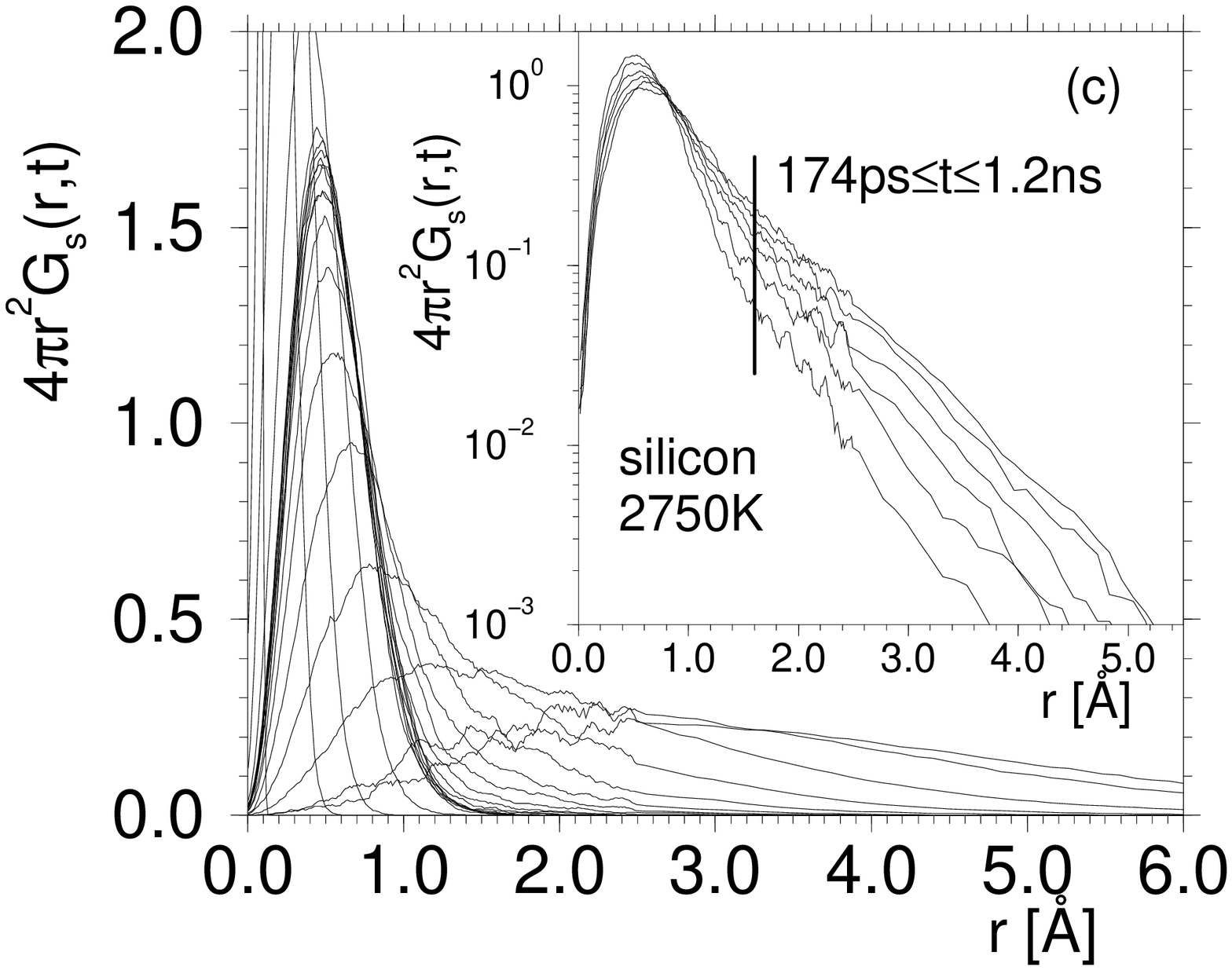,width=13cm,height=9.5cm}
\psfig{file=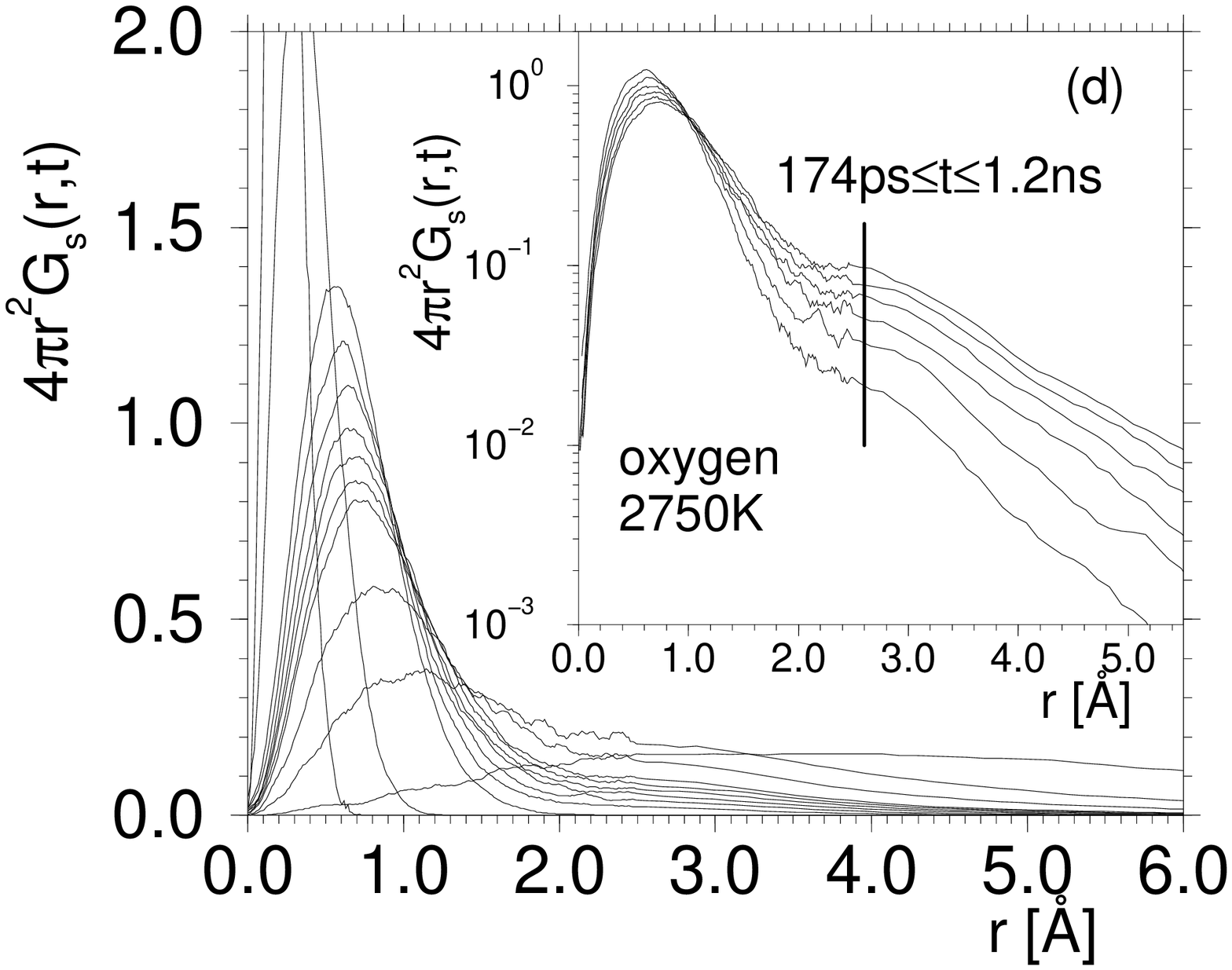,width=13cm,height=9.5cm}
\caption{Space time dependence of the self part of the van Hove
correlation function. a) silicon at 6100~K. b) oxygen at 6100~K. c)
silicon at 2750~K. The vertical line in the inset corresponds to 1.6
\AA, the length of a Si-O bond. d) oxygen at 2750~K. The vertical line
at 2.6 \AA~in the inset corresponds to the nearest neighbor O-O
distance.  In panels a) and b) the times increases by about a factor of
two from one curve to the other. In panels c) and d) the times are
0.015~ps, 0.033~ps, 0.38~ps, 174~ps, 362~ps, 590~ps, 817~ps, 1.04~ns,
1.2~ns, 4.5~ns, and 15.4~ns.}
\label{fig13}
\end{figure}

A more detailed picture of the change of the transport mechanism can
be obtained by investigating the self part of the van Hove
correlation function $G_s(r,t)$~\cite{boon_yip80} which is defined by
\begin{equation}
G_s^{\alpha}(r,t)=\frac{1}{N_{\alpha}}\sum_{i=1}^{N_\alpha}
\langle \delta(r-|{\bf r}_i(t)-{\bf r}_i(0)|) \rangle \qquad 
\alpha \in \{{\rm Si,O}\} \quad .
\label{eq13}
\end{equation}
Thus $4\pi r^2G_s^{\alpha}(r,t)$ is the probability to find a particle
at time $t$ a distance $r$ away from the place it was at $t=0$. In
Fig.~\ref{fig13} we show this probability for different times for
$T=6100$~K and $T=2750$~K. From the figures at the higher temperature
we see that the space and time dependence of $4\pi r^2G_s^{\alpha}$ is
very regular in that with increasing time the location of the peak
moves continuously to larger distances. At this temperature no
significant qualitative difference between the curves for the silicon
and oxygen atoms is observed thus showing that the transport mechanism
of the two types of particles is very similar. 

At low temperatures the situation is quite different. After the
ballistic motion of the ions, during which the $r$ and $t$ dependence
of $G_s^{\alpha}$ is qualitatively similar to the one at high
temperatures, the distribution function shows a peak whose location
depends only weakly on time.  Such a behavior is well known from
studies of the dynamics of supercooled liquids and is a manifestation
of the cage effect discussed above~\cite{roux89,kob95b}.  With
increasing time the height of this peak decreases and to the left of
the peak a long tail is observed in the case of silicon and a small
peak in the case of oxygen (around 2.6~\AA). This peak can be seen
better in a lin-log representation of the curves, which is shown in the
insets of the figures. In their simulation of a soft-sphere system Roux
{\it et al.}~\cite{roux89} related the existence of such a peak to the
presence of hopping processes and thus we have direct evidence that
such processes exist in the present system as well. Also included in
the figure for the oxygen is the location of the first peak in the
radial distribution function of the O-O correlation at 2.6\AA~(vertical
line), and we see that this location coincides with the location of the
secondary peak in $4\pi r^2G_s^{\rm O}(r,t)$.  (We also mention that
most of the oxygen atoms jumping are defects in that they are only one-
or three-fold coordinated~\cite{horbach_diss}.)

For silicon the situation is different in that no secondary peak is
visible at any time. However, a careful inspection of the curves
reveals that at around 1.6\AA, marked in the inset by a vertical line, a
change in the slope of the curves can be noticed. This distance
corresponds to the Si-O bond in one tetrahedron. From
Fig.~\ref{fig3}a we see that there are still quite a few five-fold
coordinated silicon atoms in the melt and we have found that at low
temperatures the transport of the particles often involves such
five-fold coordinated silicon atoms~\cite{horbach_diss}. Thus a typical
bond breaking process involves that a one-fold coordinated oxygen atom
attaches itself to a four-fold coordinated silicon atom, thus moving
the latter by a distance of the order of half the Si-O bond-length.
Shortly after this, the now five-fold coordinated silicon atom breaks
the bond with one of its five nearest neighbors, thus creating a new
one-fold coordinated oxygen atom, and moves again on the order of half
a bond-length. Thus in this whole process the silicon atom has moved
on the order of the length of a Si-O bond and hence $4\pi
r^2G_s^{\rm Si}(r,t)$ shows at this distance the mentioned feature.

\begin{figure}[h]
\psfig{file=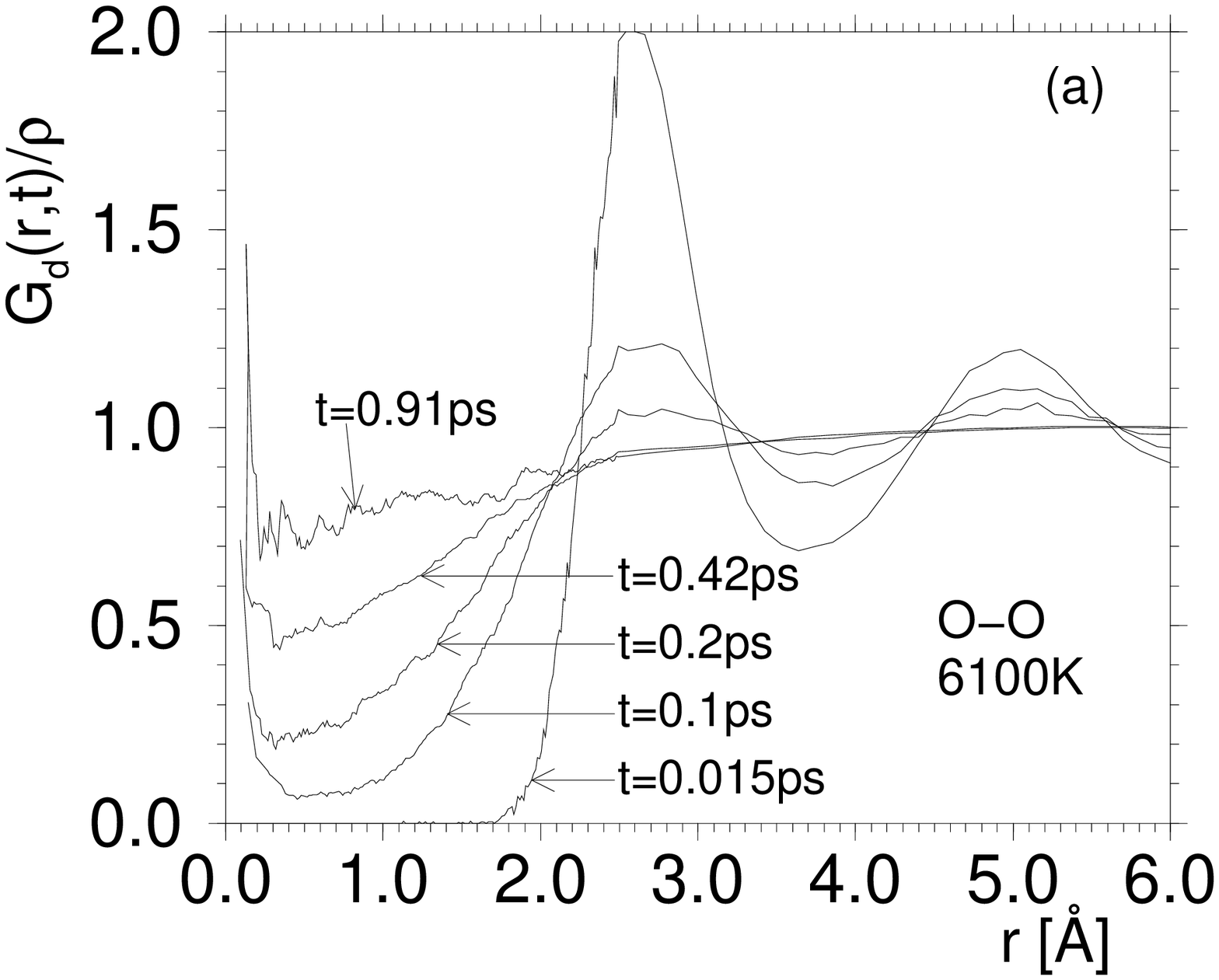,width=13cm,height=9.5cm}
\psfig{file=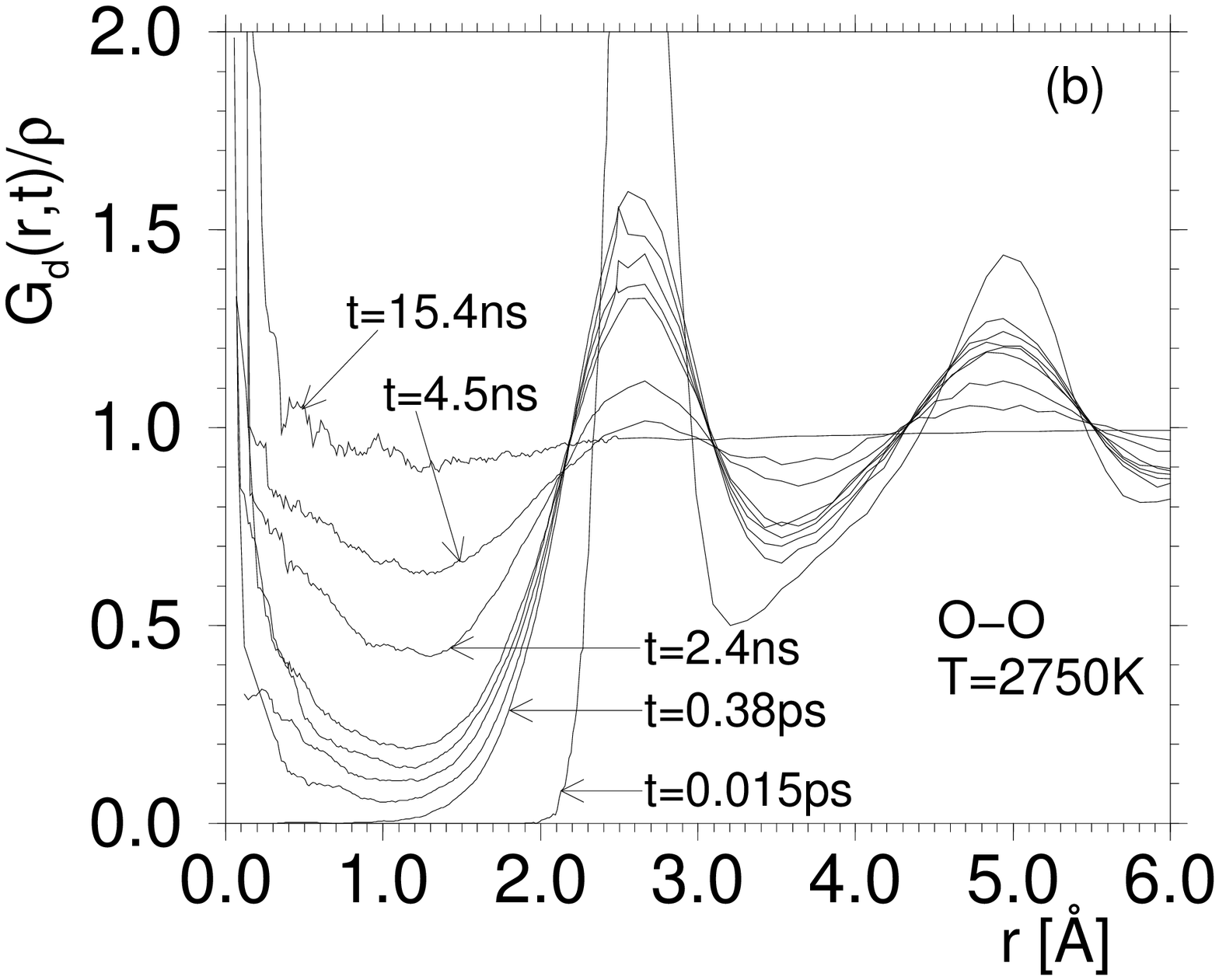,width=13cm,height=9.5cm}
\caption{Space and time dependence of the distinct part of the van
Hove correlation function for the O-O pair. a) $T=6100$~K. b)
$T=2750$~K. In panel b) the curves that are not labeled correspond to
times 174~ps, 362~ps, 590~ps, and 1.2~ns.} 
\label{fig14}
\end{figure}

Finally we mention that we have also investigated the space and time
dependence of the distinct part of the van Hove correlation function,
$G_d(r,t)$~\cite{kob95b,boon_yip80} and found that this dependence is
compatible with our discussion of the self part $G_s^{\alpha}(r,t)$.
This is demonstrated in Fig.~\ref{fig14} where we show $G_d(r,t)$ for
the O-O pair at 6100~K and 2750~K. From Fig.~\ref{fig14}a we see that
even at $T=6100$~K $G_d(r,t)$ shows at intermediate times a
small peak at the origin, $r=0$, thus showing that the correlation hole
that is observed at time zero is filled up not only by a continuous
influx from other oxygen atoms but to some extent also by particles
which hop directly into the position of the oxygen atom which was there
at time zero.  This type of dynamics is not observed in simple liquids
at high temperatures~\cite{ss_sim,roux89,kob_lj} and is thus probably
attributable to the fact that silica forms a well defined network even
at high temperatures thus making the correlation hole relatively stable
and hence permitting that a new oxygen atom can jump into it.  We
emphasize, however, that at these high temperatures this jump mechanism
is not the dominant transport mechanism, since the self part of the van
Hove function does not show a secondary peak (see Fig.~\ref{fig13}b).

In Fig.~\ref{fig14}b we see that at low temperatures the peak at the
origin is very pronounced at intermediate times and is even observable
at $t=15.4$~ns, i.e. at times at which the particles show a diffusive
behavior (see Fig.~\ref{fig7}b).  This shows that even on these very long
time scales the local structure of the network still has some
resemblance to the one at time zero and that the motion of the atoms
is indeed given by the hopping of the atoms in this structure.

\section{Summary and Conclusions}
\label{sec4}
In this paper we have discussed the results of a large scale molecular
dynamics computer simulation of a silica melt. The potential used is
the one proposed by van Beest {\it et al.}~\cite{beest90} which has
been found in previous simulations~\cite{vollmayr96b} to be quite
reliable with respect of reproducing structural properties of amorphous
silica at temperatures below the glass transition temperature. In the
present work we focus on the structural and dynamical quantities in the
{\it equilibrium} melt. Apart from thermodynamic quantities like the
pressure and the compressibility, we have determined the partial
structure factors. From these functions we conclude that the melt is
substantially ordered even at temperatures as high as 6100~K and shows
intermediate range order (first sharp diffraction peak) already at
around 4000~K. Evidence that this high degree of local ordering is not
just an artifact of our model is given firstly by the fact that at
these temperatures the viscosity of real silica is surprisingly
high~\cite{urbain82}, thus showing that the motion of the atoms is
strongly hindered because of the cage-effect, and secondly by our
finding that, if we cool the system to a temperature at which the
structure can be measured, we find that the structure predicted by the
simulation agrees very well with the experimental neutron scattering
function $S_n(q)$, thus showing that the structure of the model is very
realistic.

From the mean squared displacement of the particles we calculate the
diffusion constants $D$ and find that at low temperatures they show an
Arrhenius dependence with activation energies which agree very well
with the ones found in experiments. If this temperature dependence is
extrapolated to values of $D$ which correspond to the diffusion
constants of real silica around the {\it experimental} glass transition
temperature $T_g$, we find that these values of $D$ are at temperatures
which agree with $T_g$ to within 10\%. A similar result is
obtained from the viscosity data. Thus we conclude that the model
used is very reliable to predict the relaxation behavior of real silica
and is even able to predict the real glass transition temperature with
surprising accuracy.

For temperatures higher than 3200~K we find significant deviations from
the Arrhenius dependence of $D$. In that temperature range the
diffusion constants are fitted much better by a power-law, $D\propto
(T-T_c)^{\gamma}$, a temperature dependence which is often found in
simple liquids and which has been proposed by mode-coupling theory
(MCT). The critical temperature $T_c$ is 3330~K, in excellent agreement
with extrapolations by Hess {\it et al.} of experimental
data~\cite{hess96}. A similar conclusion can be drawn from the
temperature dependence of the viscosity. Thus we find that the
relaxation dynamics of this system shows a cross-over from a dynamics
at high temperatures which can be described by the ideal version of MCT
to an activated dynamics at low temperatures. The existence of this
cross-over can also be seen in the temperature dependence of the
product of the diffusion constants and the life time of a Si-O bond.
This product is constant in the high temperature regime for silicon and
oxygen, whereas it shows a significant (Arrhenius) temperature
dependence for silicon at low temperatures, hence giving evidence for a
change of the diffusion mechanism with decreasing temperature.

As it has been shown elsewhere~\cite{kob98b}, even the activated
dynamics at low temperatures can be understood to a good part within
the framework of MCT, since predictions of the theory like the
factorization property, or the existence of the von Schweidler
law~\cite{mct} seem to hold very well. Thus we come to the conclusion
that one of the main differences between strong and fragile glass
formers is that in the former the so-called hopping processes, which
invalidate some of the predictions of the {\it ideal} version of MCT,
are, {\it at} $T_c$, very pronounced in strong glass formers. Due to
the strong presence of these hopping processes important predictions of
the ideal MCT, like the presence of a power-law dependence of the
transport coefficients, are valid only in a very restricted temperature
range. Therefore it is important that in the analysis of the relaxation
dynamics of strong glass formers those predictions of MCT are checked
which are not affected by the presence of such hopping processes, if
one wants to make a real test whether or not MCT is able to describe
the relaxation dynamics of such a system.

Acknowledgements: We thank K. Binder for many stimulating discussions
on this work, J.-L. Barrat for pointing out
Refs.~\cite{allen94,visc_ewald} to us, and J. Baschnagel for useful
suggestions. This work was supported by BMBF Project 03~N~8008~C and by
SFB 262/D1 of the Deutsche Forschungsgemeinschaft.  We also thank the
HLRZ J\"ulich for a generous grant of computer time on the T3E.

\end{document}